%% file: main.tex
\newif\ifpreprint
\preprinttrue

\documentclass[ProofPDF,WebPDF]{./Styles/Jnls-Template-V3}

\arrayrulecolor{BlueTextColor}

\gdef\printartdetails{}

\rectoheadtop{\RHAuthor}
\versoheadtop{\RHAuthor}

\ifpreprint\else
  \AtBeginDocument{\hypersetup{pdfauthor={}, pdfsubject={}, pdfkeywords={}}}
\fi

\newcommand{\suppsec}[2]{Supplementary File, Appendix~#2}
\newcommand{\supptab}[2]{Supplementary File, Table~#2}

\newcommand{\mainsec}[2]{Section~#2 of the main paper}

\usepackage{adjustbox}

\begin{document}

\jname{Transactions of the International Society for Music Information Retrieval }
\jvol{00}
\jissue{00}
\jmonth{season}
\jyear{xxx}
\jid{doi}
\artid{xx.xxxx/xxxx.xx}
\filename{main}
\def\MSFirstPage{\pageref{startPage_Label}}
\makeatletter
% \MSLastPage is expanded in the running head/foot while a page is shipped
% out, long before \label{endPage_Label} at the end of the document is read.
% A bare \pageref there emits "Reference undefined", which arXiv treats as a
% blocking error. Guard it: print the number once the label is in the .aux,
% print nothing on the pass before that. The final PDF is unchanged.
\def\MSLastPage{\ifcsname r@endPage_Label\endcsname\pageref{endPage_Label}\fi}
\makeatother
\graphicspath{{./figures/}}
\label{startPage_Label}
\title{A Dual Evaluation for Music Transcription}

\articletype{RESEARCH ARTICLE}

\ifpreprint
  \author[1]{Ping Wang}
  \howtociteauthor[1]{Ping Wang}
  \rhauthor[1]{P. Wang}
  \author[2]{Guang Yang}
  \howtociteauthor[2]{Guang Yang}
  \rhauthor[2]{G. Yang}
  \author[3]{Nazif Can Tamer}
  \howtociteauthor[3]{Nazif Can Tamer}
  \rhauthor[3]{N. C. Tamer}
  \author[4]{Victoria Ebert}
  \howtociteauthor[4]{Victoria Ebert}
  \rhauthor[4]{V. Ebert}
  \author[5]{Noah A. Smith}
  \howtociteauthor[5]{Noah A. Smith}
  \rhauthor[5]{N. A. Smith}

  \aff[1]{University of Washington, Seattle, WA, USA, \href{mailto:pingw220@cs.washington.edu}{pingw220@cs.washington.edu}}
  \aff[2]{University of Washington, Seattle, WA, USA, \href{mailto:gyang1@cs.washington.edu}{gyang1@cs.washington.edu}}
  \aff[3]{University of Washington, Seattle, WA, USA, \href{mailto:nctamer@cs.washington.edu}{nctamer@cs.washington.edu}}
  \aff[4]{University of Washington, Seattle, WA, USA, \href{mailto:ebertv@cs.washington.edu}{ebertv@cs.washington.edu}}
  \aff[5]{University of Washington and Allen Institute for Artificial Intelligence, Seattle, WA, USA, \href{mailto:nasmith@cs.washington.edu}{nasmith@cs.washington.edu}}
  % The banner lives here because \afflnote is the only note slot the title
  % block actually typesets: Jnls-Template-V3.cls:2906 emits it after a \vfill
  % inside the title \vbox, i.e. at the foot of the title panel on page 1.
  % \subtitle is defined by the class but the \@maketitle this template
  % resolves to (cls:2838) never expands it, so a \subtitle would vanish
  % silently. %TC:ignore keeps the banner out of the 8,000-word count.
  %TC:ignore
  \afflnote{{\fontsize{9bp}{11bp}\selectfont\bfseries Preprint. Under review at
  \emph{Transactions of the International Society for Music Information
  Retrieval} (TISMIR).}\\
  *Author affiliations can be found in the back matter of this article}
  %TC:endignore
\else
  \author[1]{Anonymous Author(s)}
  \howtociteauthor[1]{Anonymous Author(s)}
  \rhauthor[1]{Anonymous}
  \aff[1]{Affiliation withheld for double-blind review}
\fi

\begin{abstract}[ABSTRACT]
Automatic music transcription systems produce sheet music that can be read and played back.  We argue that these two targets call for complementary evaluations of notation similarity to a reference score and playback similarity to the original performance, respectively. Our study considers notation similarity metrics from the optical music recognition literature and a wide range of playback-similarity methods validated through a listening study across over 100 participants and 230 piano recordings covering 23 works, 30 performers, and six composers.  We find, fortuitously, that the playback similarity metric that correlates best with human judgments, CLEWS, is also the cheapest to run.  We also find that the two evaluation dimensions favor different systems among a collection of 24 pipelines formed by pairing eight audio-to-MIDI models with three MIDI-to-score converters, with the latter component systematically determining the favored objective. The complementarity between metrics also holds when adding to the pool Rubato, a new end-to-end system that offers substantially improved notation similarity while remaining competitive, though not the best, on playback similarity.
\end{abstract}

\begin{keyword}[KEYWORDS:]

automatic music transcription, dual evaluation, evaluation methodology, human listening study, notation similarity, playback similarity

\end{keyword}

\maketitle

\ifpreprint\else\linenumbers\fi

\section{Introduction}

Music is primarily experienced through two principal modalities: sight and sound. For many musical genres, engraved musical notation (sheet music) communicates performance instructions through spatial layout and notation structure. When performers play back the sheet music, they render the symbolic notation into sound experienced by the listener. Importantly, no two musicians will play the same sheet music the same way. Differences in tempo, dynamics, and timing will be audible but not visible in conventional visual notation. If a piece of sheet music were always played identically by every performer, there would be no need to ever record versions other than the original. If sheet music encoded every difference between performances, it would likely be unreadable by musicians.

In this work, we evaluate the tradeoff in notational and playback fidelity of 25 complete automatic music transcription (AMT) systems --- 24 modular pipelines and one end-to-end model --- and highlight the importance of a multi-view evaluation of AMT.

AMT has been decomposed into subtasks, including beat tracking \citep{foscarin2024beat}, audio-to-MIDI note detection \citep{bradshaw2024musically, 2022_BittnerBRME_LightweightNoteTranscription_ICASSP, kong2020high, gardner2022mt3, yan2021skipping, yan2024scoring, hawthorne2021sequence, simon2022scaling}, and MIDI-to-score conversion \citep{cuthbert2010music21, musescore, beyer2024end}. These subtasks emphasize different parts of the audio-to-score pipeline: audio-to-MIDI systems focus on recovering performed musical events from sound, while MIDI-to-score systems focus on converting symbolic events into readable notation. However, evaluating a complete audio-to-score system, as opposed to a component, requires considering both how the output appears as a written musical document and how it sounds when rendered back to audio.

\begin{figure*}[t]
    \centering
    \includegraphics[width=\textwidth]{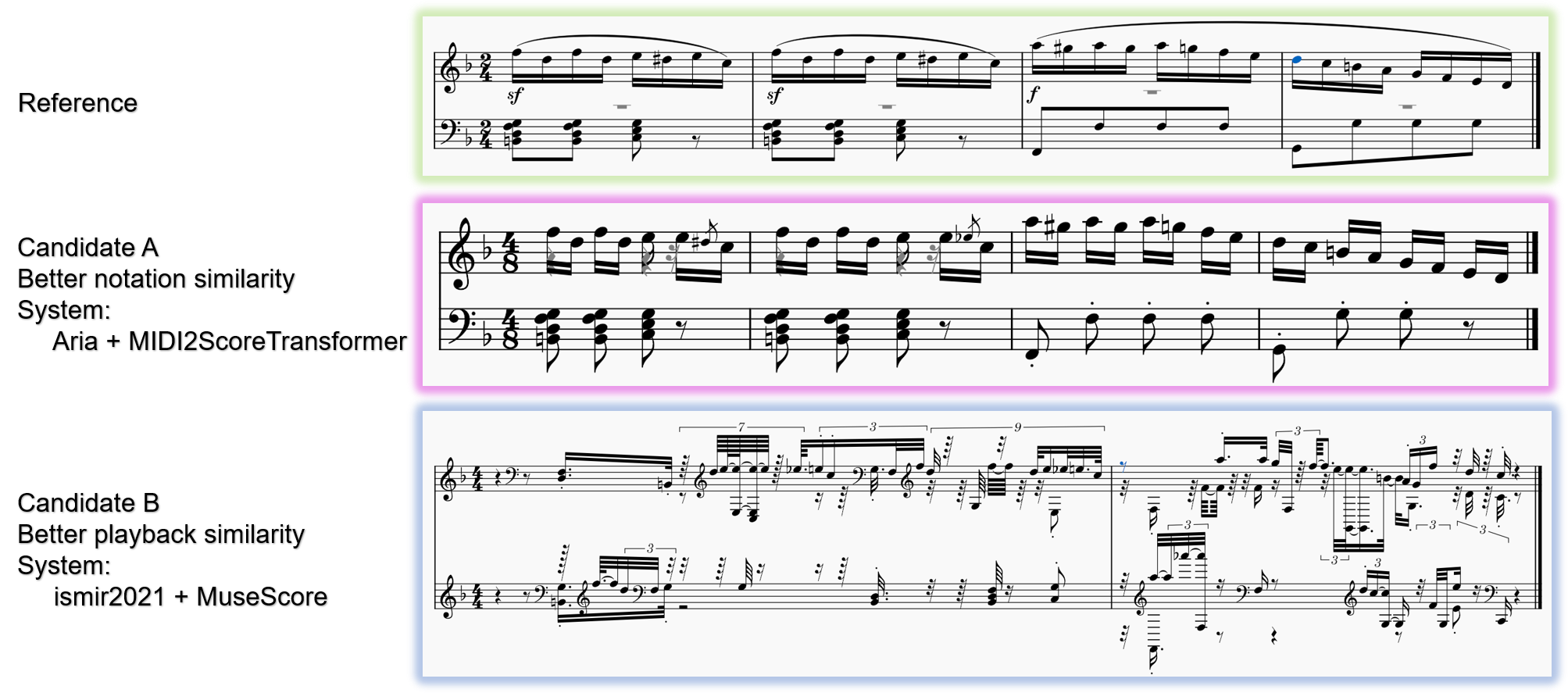}
    \caption{Tradeoff between notation and playback similarity in audio-to-score
    transcription. The top row is a four-measure reference excerpt from Beethoven,
    \textit{Piano Sonata No.~6}, third movement, Op.~10 No.~2. Candidate A is
    produced by Aria-AMT~\citep{bradshaw2024musically} + MIDI2ScoreTransformer~\citep{beyer2024end} and Candidate B by
    ismir2021~\citep{hawthorne2021sequence} + MuseScore~\citep{musescore}. Candidate A is more similar to the reference as written notation (OMR-NED $=49.3$) but less similar in rendered playback (TWED cost $=1.260$); Candidate B is the opposite (OMR-NED $=98.1$, TWED cost $=0.322$). Lower OMR-NED and lower TWED are better. Notation-side and playback-side evaluation favor different outputs.
    }
    \label{fig:intro-tradeoff-example}
\end{figure*}

We therefore formulate audio-to-score transcription evaluation as a two-dimensional problem. The first dimension seeks \emph{notation similarity}:  the generated sheet music should closely match the reference sheet music as a written musical document. The second dimension seeks \emph{playback similarity}:  audio rendered from the generated sheet music should faithfully reproduce the music in the original performance recording. These dimensions are complementary: transcribed sheet music with high notation similarity may still have low playback similarity when rendered (e.g., because of an expressive performance), while a transcription with high playback similarity may offer a symbolic representation with low notation similarity or poor readability. The examples in Figure~\ref{fig:intro-tradeoff-example} illustrate this tension.

To study this tradeoff between notational similarity and playback similarity, we evaluate 24 complete audio-to-score transcription pipelines formed by combining eight competitive and widely used audio-to-MIDI models with three MIDI-to-score converters. These components include recent transformer-based systems, piano-specific transcription models, lightweight transcription baselines, and commonly used notation-conversion tools, giving broad coverage of strongest publicly available audio-to-score pipelines. For notation similarity, we compare generated MusicXML scores against reference scores using \textbf{OMR Normalized Edit Distance (OMR-NED; \citealp{martinezsevilla2025smb})}, a normalized edit-distance metric designed for symbolic music notation comparison. For playback similarity, we render each generated score back to audio and compare it with the original recording. This playback-side evaluation includes digital signal processing (DSP)-based feature-alignment metrics, \textbf{dynamic time warping~(DTW)}~\citep{sakoe1978dynamic} and \textbf{time warp edit distance~(TWED)}~\citep{marteau2009time}, applied to multiple frame-level audio representations spanning \emph{pitch-class and harmonic}, \emph{spectral}, and \emph{rhythm and onset} information, thereby capturing complementary aspects of pitch, harmony, timbre, onset activity, tempo, and temporal structure; \textbf{human ABX listening judgments}; pairwise \textbf{AudioLM} judges; and \textbf{learned embedding-similarity evaluators}.

By considering notational fidelity as a separate dimension from playback fidelity, we cast light on the practical tradeoff between them: different transcription systems, and the developers who build them, may reasonably choose to optimize for different uses, from producing clean, musician-readable scores to capturing the details needed for faithful audio resynthesis or performance analysis.

Our contributions are:
\begin{itemize}
    \item We formulate audio-to-score transcription evaluation as a two-dimensional problem seeking both notation and playback similarity.
    \item We evaluate 24 complete audio-to-score pipelines that produce MusicXML scores from audio recordings, separating the effects of eight audio-to-MIDI models and three MIDI-to-score converters.
    \item We collect a human ABX listening reference with 106 participants and 3,180 pairwise judgments, and use it to validate DTW, TWED, and AudioLM judges, and learned embedding evaluators, including CLaMP~3 and CLEWS.
    \item We show that notation-side and playback-side evaluation can prefer different converter families, motivating multi-view evaluation and tradeoffs in system design.
    \item We evaluate a contemporaneously developed end-to-end transcription system, Rubato \citep{tamer2026rubatotranscribingpianomusic} in the same framework.
\end{itemize}

\section{Methodology}
\label{sec:methodology}

Our proposed evaluation setup distinguishes two primary dimensions: \emph{notation similarity}~(\S\ref{subsec:notation_similarity}) and \emph{playback similarity}~(\S\ref{subsec:playback_similarity}).

\subsection{Notation similarity}
\label{subsec:notation_similarity}

Our notation-side evaluation treats the generated sheet music as a \emph{visual, written musical document}. We use \textbf{OMR Normalized Edit Distance (OMR-NED)}~\citep{martinezsevilla2025smb}, a metric developed for optical music recognition (OMR) evaluation. OMR-NED compares explicit notation objects in the predicted and reference sheet music. %
Although its input here is a symbolic score rather than a page image, its target is the visual notation domain: it asks whether the symbols that would appear on the page match the reference notation. OMR-NED is hence the best available operationalization of the kind of notation similarity sought by AMT; see \S\ref{subsec:related-amt-evaluation} for a discussion of alternatives.

OMR-NED defines edit operations over music-notation symbols: note and rest properties such as pitch, accidentals, note heads, beams and ties, and non-note objects such as clefs, key and time signatures, dynamics, slurs and lyrics. It therefore measures differences in the sheet music that would require manual correction by a user of an OMR (or AMT) system.

Let $\hat{Y}$ denote a transcription and $Y$ the reference sheet music, and let $I(\hat{Y},Y)$ and $D(\hat{Y},Y)$ be the numbers of symbol insertions and deletions in the cheapest transformation of $\hat{Y}$ into $Y$. The per-score OMR-NED is
\begin{equation}
\mathrm{OMR\text{-}NED}(\hat{Y},Y)
=
\frac{I(\hat{Y},Y) + D(\hat{Y},Y)}
{|\hat{Y}| + |Y|},
\end{equation}
where $|\cdot|$ counts notation symbols, so the metric is length-aware. Lower is better and zero is an exact match. Following \citet{martinezsevilla2025smb}, we report the micro-average over evaluated files,
\begin{equation}
\mathrm{Overall\ OMR\text{-}NED}
=
\frac{
\sum_k I(\hat{Y}_k,Y_k) + \sum_k D(\hat{Y}_k,Y_k)
}{
\sum_k |\hat{Y}_k|+ \sum_k |Y_k|
},
\end{equation}
where $k$ indexes evaluated recording and sheet-music pairs. For readability, all reported OMR-NED values, in both tables and figures, are scaled by $100$. The full symbol inventory and the treatment of substitutions are given in \suppsec{app:omr-ned-details}{A.1}.

\subsubsection{Related work: AMT evaluation}
\label{subsec:related-amt-evaluation}

Most automatic music transcription evaluation is reference-based: a system output is compared with a ground-truth representation, and the metric defines which differences matter. For audio-to-MIDI transcription, this comparison is usually made over frame-level pitch activations or note events, using precision, recall, and F-measure over estimated pitches, onsets, and offsets~\citep{raffel2014mireval}. These metrics are appropriate when the target output is MIDI-like note events. However, for transcription that produces sheet music as its final output, evaluation must consider whether the resulting notation is musically interpretable and visually usable.

Three notation-oriented metrics are the closest alternatives to OMR-NED. \citet{cogliati2017metric} count notation errors across dimensions such as notes, durations, rests, barlines and staff assignment, and map them to human ratings with a regression fitted on a small rated set; the counts are not disjoint, so one structural mistake can be penalised several times. \textbf{MV2H}~\citep{mcleod2018mv2h,mcleod2019nonaligned} enforces disjoint penalties over five dimensions, but targets musical structure rather than engraving, and its MusicXML workflow compares MIDI-derived events rather than notation objects. \textbf{MUSTER}~\citep{nakamura2018towards,hiramatsu2021joint} reports separate error rates, of which only five note-oriented ones enter its MeanER.

We applied MUSTER to the same sheet music. Its subscores do not support a consistent interpretation (\supptab{tab:appendix-muster-correlations}{A.1}): the structure-oriented VoiceER, HandER and ScaleErr track OMR-NED ($\rho=0.94$, $0.81$, $0.81$), whereas MeanER and OnsetER track human playback preference far more closely ($\rho=0.79$ and $0.84$) than OMR-NED ($\rho=0.38$ and $0.50$). Choosing a different subscore therefore changes the conclusion, while MeanER omits score-organization properties central to the visual task. We use OMR-NED for the main notation comparison; full MUSTER scores and ranks are in Supplementary File, Tables A.4 and~A.5, and the three metrics are discussed at length in \suppsec{app:extended-related-work}{A.3}.

\subsection{Playback similarity}
\label{subsec:playback_similarity}

We treat playback similarity as the second evaluation dimension measured through multiple evaluator families. We first render each transcribed score to audio (\S\ref{subsubsec:audio_rendering}) and extract frame-level audio representations from the original and rendered recordings (\S\ref{subsubsec:audio-features}). We then compare the recordings using deterministic DSP-based feature-alignment metrics, DTW and TWED (\S\ref{subsubsec:feature-alignment-metrics}), and learned or model-based evaluators, including AudioLM judges, CLaMP~3, and CLEWS (\S\ref{subsubsec:model-based-metrics}). We validate these automatic evaluators against human pairwise listening judgments in \S\ref{subsec:human-validation} and compare their practical costs in \S\ref{subsec:evaluation-cost}.

\subsubsection{Audio rendering}
\label{subsubsec:audio_rendering}

Audio-based evaluation requires converting symbolic representation into audio. Therefore, for each system, we convert the predicted MusicXML score to MIDI and then to waveform audio. We use the same rendering setup across systems.

The rendering pipeline uses MuseScore to export MusicXML files to MIDI, followed by FluidSynth~\citep{fluidsynth} to synthesize WAV audio using the SGM soundfont~\citep{sgm}. MuseScore is used here only as a fixed downstream renderer, and its relevant import/export routines are rule-based rather than a shared learned bidirectional model; we therefore do not expect that using MuseScore's rendering abilities in this way gives any advantage to evaluated pipelines that use MuseScore for MIDI-to-score conversion. The rendered audio is produced at $44.1$ kHz and peak-normalized.

\subsubsection{Audio feature representations}
\label{subsubsec:audio-features}
Playback similarity is computed over frame-level audio features. Because transcription quality is multidimensional---pitch, timing, rhythm, duration, and harmony---and no single feature captures all of it, we evaluate fourteen features \citep{mcfee2015librosa,bogdanov2013essentia} in three groups: \emph{pitch-class and harmonic} (Chroma, HPCP, Chroma CENS, Tonnetz, CQT, STFT semitone, pitch salience), capturing note content, harmonic structure and tonal motion; \emph{spectral} (Mel spectrogram, MFCC, PCEN-Mel, Log-STFT, spectral flux), which can be sensitive to rendering artifacts; and \emph{rhythm/onset} (onset strength, tempogram). Each feature is listed with its citation and musical property in \supptab{tab:audio-features}{A.7}.

All reference and rendered audio is processed at $44.1$~kHz. Frame-based features use a $10$~ms hop with frame-wise normalization before distances are computed, and tempogram is derived from onset strength. Each representation is scored independently: per feature we compute one reference--candidate DTW or TWED distance per recording and average over the evaluation set to obtain one system-level score. The per-feature scores are compared in \S\ref{subsec:human-validation}.

\subsubsection{Feature-alignment playback metrics}
\label{subsubsec:feature-alignment-metrics}

After extracting frame-level audio features, we compare each reference recording with the audio rendered from a candidate score using two established time-series alignment distances: \textbf{dynamic time warping~(DTW)}~\citep{sakoe1978dynamic} and \textbf{time warp edit distance~(TWED)}~\citep{marteau2009time}. We use these metrics because score renderings and real performances may differ in local tempo, timing, and duration even when they preserve similar musical content. Alignment-based distances therefore provide a more flexible comparison than framewise distances computed at fixed time indices.

\textbf{DTW} finds a minimum-cost alignment path between the reference and rendered feature sequences. We evaluate each of the 14 feature representations separately. For a given representation, each frame is represented by a feature-specific multidimensional vector, such as pitch-class bins, spectral coefficients, or multiband onset-strength values. We use cosine distance between the corresponding reference and candidate frame vectors as the local alignment cost and normalize the accumulated cost by the length of the warping path. Lower DTW cost indicates greater similarity between the two feature trajectories.

Compared with DTW, \textbf{TWED} explicitly penalizes temporal gaps and edit operations, making it a stricter alignment metric that can reduce over-alignment. TWED is parameterized by scalars $\nu$, which controls temporal stiffness, and $\lambda$, which controls the edit/deletion penalty. We evaluate four settings, $(\nu,\lambda)\in\{(0.001,1.0),(0.01,1.0),(0.001,2.0),(0.01,2.0)\}$.

Because DTW and TWED are distances (lower is better), we negate their system-level costs when computing rank correlations against higher-is-better human and model-based scores. Full definitions, local costs, and normalization are given in \suppsec{app:dtw-twed-details}{A.5}.

\subsubsection{Model-based and embedding playback metrics}
\label{subsubsec:model-based-metrics}

Beyond DSP feature alignment, we consider three families of learned playback-side evaluators, defined here and validated against the human listening study in \S\ref{subsec:human-validation}: pairwise \textbf{AudioLM} judges, \textbf{CLaMP~3} embeddings, and \textbf{CLEWS} embeddings. Unlike DTW and TWED, these are model-based and can shift with model version, prompt, embedding preprocessing, or API availability, a tradeoff we quantify in \S\ref{subsec:evaluation-cost}.

\textbf{AudioLM} evaluation uses the same reference-anchored format as the human task (explained more in \S\ref{subsec:human-abx}): the model receives reference clip $X$ and candidates $A$ and $B$ --- fixed 10-second excerpts, rather than the untruncated playback offered to human listeners --- and chooses which better matches $X$ by musical content (pitch, onset, duration), returning exactly $A$ or $B$ with no tie, explanation, or confidence rating. We evaluate Gemini 2.0 Flash, 2.0 Flash-Lite, 2.5 Flash, 2.5 Flash-Lite, 2.5 Pro, 3 Flash, 3 Pro, 3.1 Flash-Lite, and 3.1 Pro as a screening sweep, and report the strongest live model, Gemini 3.1 Pro, as the primary AudioLM judge. Outputs are converted to pairwise wins via the same side-mapping as the human responses; a response is valid if it parses as exactly $A$ or $B$, and empty/ambiguous responses are excluded. Because a single-order presentation exposes an AudioLM judge to a position bias (we observe a $\sim$59\% preference for the first candidate), we judge each pair in \emph{both} $A/B$ orders and average, so position effects cancel; the human study instead randomized order per trial. For the screening sweep each model runs $\approx 1{,}800$ single-order trials ($N=60$ blocks of 30), and the reported Gemini 3.1 Pro judge runs all $1{,}800$ pairs in both orders ($3{,}600$ trials).

\textbf{CLaMP~3}~\citep{wu2025clamp3} uses modality-specific encoders for audio, sheet music, and performance signals, and projects their outputs into a shared learned embedding space. We evaluate audio--audio, cross-modal audio--ABC, and ABC--ABC similarity as learned similarity signals. Although ABC encodes symbolic musical content, CLaMP~3 is trained to align different musical modalities within a common representation rather than to compare explicit notation objects directly. In this shared space, an ABC score can be close to an audio recording or MIDI-based performance when they express similar musical content, even though their surface representations are different. Consequently, the embedding is intended to preserve modality-invariant musical and semantic information, and may abstract away notation-specific differences such as beams, stem directions, rests, voice assignments, staff placement, clefs, and engraving structure. We therefore do not treat CLaMP~3 ABC similarity as a measure of notation fidelity. Instead, we analyze CLaMP~3 as a learned evaluator of broader musical correspondence across and within modalities.

\textbf{CLEWS}~\citep{pmlr-v267-serra25a} is a learned audio embedding-similarity evaluator that compares reference and rendered candidate audio through several aggregation variants (distance, similarity, global cosine, segment mean, segment median, best-match). We use segment-mean similarity as the representative variant because it correlates most strongly with human-derived scores (\S\ref{subsec:human-score-aggregation}).

\subsubsection{Related work: playback similarity evaluation for AMT systems}
\label{subsec:related-audio-faithfulness}

\citet{ycart2020perceptual} validate piano-transcription metrics with a pairwise listening test and learn a metric from the judgments, and \citet{simonetta2022perceptual} show that objective metrics computed from MIDI note events can correlate weakly with listeners' judgments of preserved interpretation. Both compare synthesized MIDI transcriptions against a synthesized MIDI reference; we instead compare audio rendered from final MusicXML scores against the original recordings, so our evaluation asks whether generated sheet music preserves the musical content actually heard. These studies are discussed further in \suppsec{app:extended-related-work}{A.3}.

\section{Dataset and Evaluated Systems}
\label{sec:dataset-systems}

Our proposed two-dimensional evaluation is designed to study notational and playback similarity achieved by automatic music transcription systems on data unseen during system development (\S\ref{subsec:dataset}). Because most publicly available systems do not transcribe full recordings directly into complete engraved scores, our primary evaluation constructs complete AMT pipelines by pairing competitive audio-to-MIDI models with MIDI-to-score converters (\S\ref{subsec:evaluated-systems}). In addition, we include Rubato~\citep{tamer2026rubatotranscribingpianomusic} as a separate end-to-end case study. Rubato differs from the modular systems evaluated in our main grid: it directly predicts timestamped piano notation rather than decomposing transcription into audio-to-MIDI and MIDI-to-score stages. It was completed and released after our main study was completed, so we report its results separately in \S\ref{subsec:rubato-case-study}, using the same notational and playback-side metrics where applicable, to contextualize how a recent end-to-end score transcription model behaves under our two-dimensional evaluation.

\subsection{Dataset}
\label{subsec:dataset}

We evaluate all transcription systems on a piano transcription dataset derived from ATEPP~\citep{zhang2022atepp}. ATEPP is a large collection of YouTube piano recordings, a subset of which include paired scores. From the score-available subset of ATEPP, we exclude recordings whose symbolic scores are sourced from ASAP~\citep{asap-dataset} in order to conservatively reduce possible overlap with training or evaluation material used by prior transcription systems. ATEPP provides the fairest comparison across models: it is the largest set unseen by all baselines, and includes substantial variability in recording conditions, including historical performances.

The main evaluation set contains 230 unique reference recordings, each with one distinct YouTube recording. Each recording is paired with an original audio file and a ground-truth symbolic score. These recordings cover 23 unique musical works by six composers and 30 distinct performers. We evaluate recordings capped at three minutes for computational efficiency.

\subsection{Evaluated systems}
\label{subsec:evaluated-systems}

We evaluate 24 pipelines formed by pairing eight audio-to-MIDI models with three MIDI-to-score converters. The audio-to-MIDI models include MT3 \citep{gardner2022mt3}, ISMIR 2021 (I21; \citealp{hawthorne2021sequence}), ISMIR 2022 small/base (I22s/I22b; \citealp{simon2022scaling}), Basic Pitch (BP; \citealp{2022_BittnerBRME_LightweightNoteTranscription_ICASSP}), ByteDance Piano (Byte; \citealp{kong2020high}), Transkun (Tkun; \citealp{yan2021skipping,yan2024scoring}), and Aria-AMT (Aria; \citealp{bradshaw2024musically}). The MIDI-to-score converters are music21 (M21; \citealp{cuthbert2010music21}), MuseScore (MS; \citealp{musescore}), and MIDI2ScoreTransformer (M2ST; \citealp{beyer2024end}). We treat each audio-to-MIDI and MIDI-to-score combination as one complete transcription system, denoted by model+converter; for example, Aria+MS refers to Aria-AMT followed by MuseScore conversion. These abbreviations are used throughout the paper.

\section{Results}
\label{sec:results}

Our results examine the relationship between preserving the reference as written notation and preserving the original performance in rendered playback. We first compare these two dimensions directly (\S\ref{subsec:main-results}) and then analyze the effects of the audio-to-MIDI model and MIDI-to-score converter (\S\ref{subsec:pipeline-component-effects}). We use OMR-NED for notation similarity and CLEWS as the representative automatic playback metric in the main figures because CLEWS achieves the strongest agreement with human preference and the best cost--agreement balance in our benchmark (\S\ref{subsec:human-validation}; \S\ref{subsec:evaluation-cost}).

The main comparison covers the 24 modular audio-to-score pipelines formed by pairing eight audio-to-MIDI models with three MIDI-to-score converters. We additionally evaluate Rubato~\citep{tamer2026rubatotranscribingpianomusic} as a case study in applying the dual evaluation methodology to a recently-arrived end-to-end system (\S\ref{subsec:rubato-case-study}). The human validation covers only the original 24 modular pipelines because Rubato was released after the listening study completed. Complete system-level scores and rankings are reported in \suppsec{app:supplementary-system-tables}{A.2}.

\subsection{Notation--playback similarity tradeoff}
\label{subsec:main-results}

\begin{figure*}[t]
\centering
\includegraphics[width=0.8\textwidth]{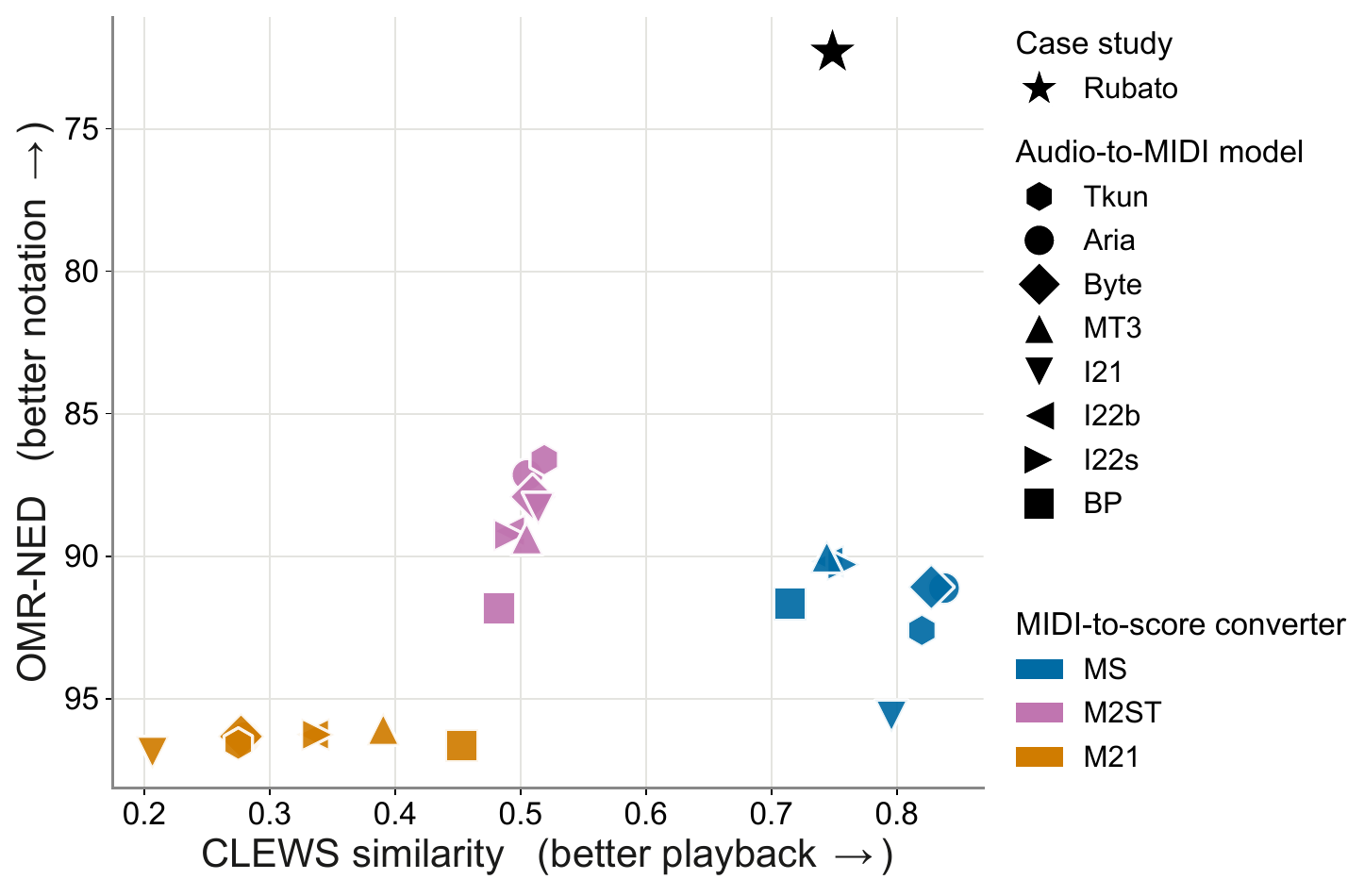}
\caption{Notation similarity versus playback similarity for the evaluated audio-to-score systems. The vertical axis is OMR-NED, inverted so that better notation similarity is upward. The horizontal axis is CLEWS similarity, with better playback similarity to the right. Marker shape identifies the audio-to-MIDI model and marker colour the MIDI-to-score converter. Rubato, shown with a black star, is keyed separately as a case study: it is end-to-end and has neither a MIDI stage nor a converter, so it belongs to neither group.}
\label{fig:omr-versus-playback-main}
\end{figure*}

Figure~\ref{fig:omr-versus-playback-main} summarizes the central result of the two-dimensional evaluation. The vertical axis reports notation similarity through OMR-NED, while the horizontal axis reports playback similarity through CLEWS. Both axes are oriented so that better systems appear upward and to the right. The 24 modular systems are formed by combining eight audio-to-MIDI models with three MIDI-to-score converters.

The converter families occupy distinct regions of the two-dimensional space. M2ST outputs form the strongest notation-side cluster, whereas MS outputs form the strongest playback-side cluster. M21 pipelines trail both families across the two views. Thus, notation and playback evaluation favor different downstream conversion strategies. We examine the component-level structure behind this separation next.

\subsection{System analysis}
\label{subsec:pipeline-component-effects}

We next examine how the audio-to-MIDI model and MIDI-to-score converter affect complete-system performance. We analyze notation similarity in \S\ref{subsubsec:notation-system-analysis} and playback similarity in \S\ref{subsubsec:playback-system-analysis}. (Rubato is discussed separately in \S\ref{subsec:rubato-case-study}.)

\subsubsection{Notation similarity}
\label{subsubsec:notation-system-analysis}

\begin{figure*}[t]
\centering
\includegraphics[width=0.8\textwidth]{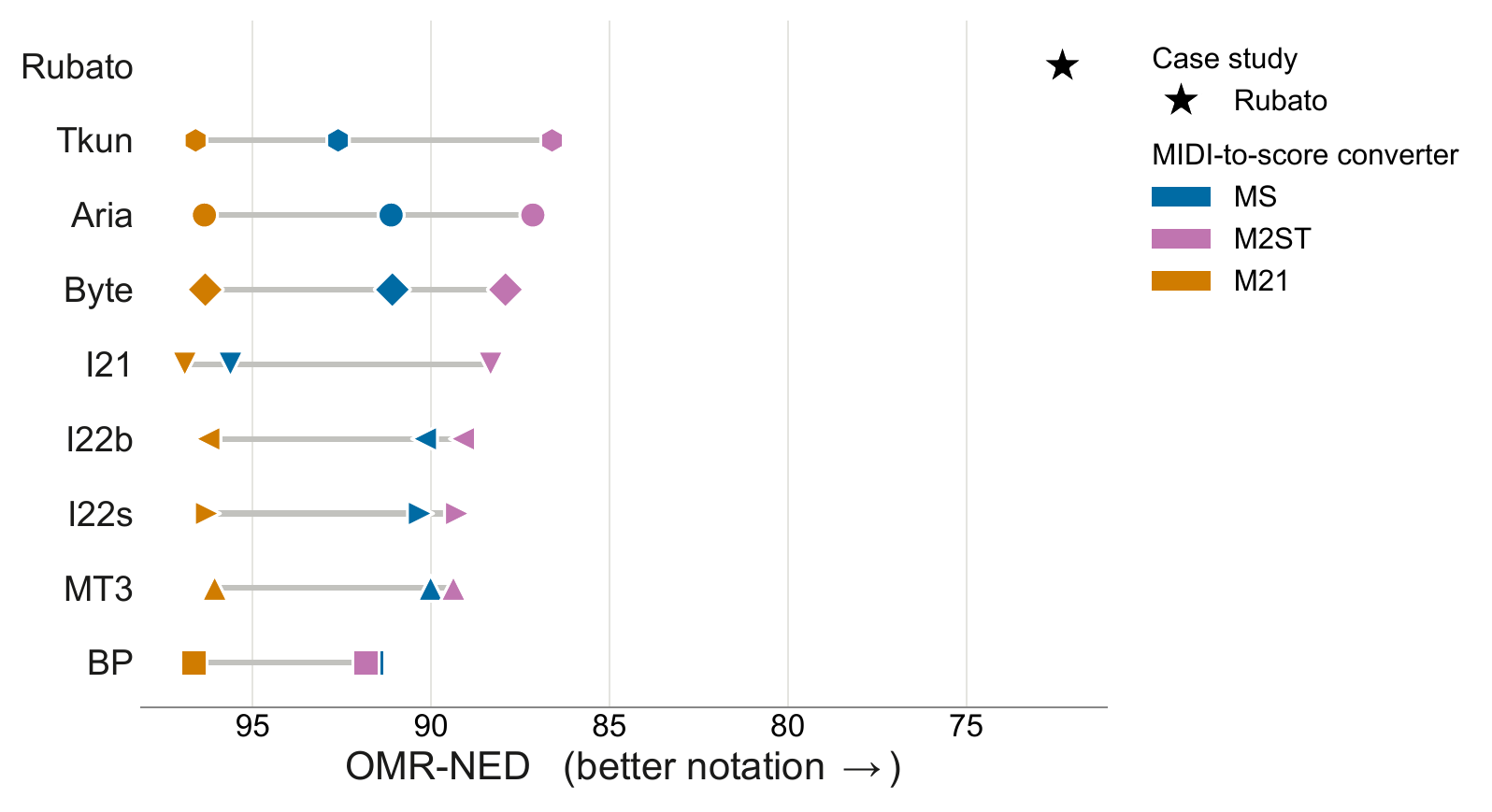}
\caption{Notation similarity across audio-to-score pipelines. Rows and marker shapes identify the audio-to-MIDI model, while marker colors identify the MIDI-to-score converter. The gray segment connects the three converter results for each transcriber. OMR-NED is lower-is-better, and the axis is reversed so that stronger notation similarity appears to the right. Rows are ordered by the rightmost point in each row, that is, by each model's best converter, strongest at the top. Rubato is shown separately as a case study, with a black star and its own legend key.}
\label{fig:omr-system-analysis}
\end{figure*}

Figure~\ref{fig:omr-system-analysis} shows a consistent converter ordering under OMR-NED. M2ST is generally the strongest notation-side converter, with Tkun+M2ST the strongest modular pipeline (OMR-NED = 86.61). MS is intermediate, whereas M21 is the weakest converter in every row.

The figure also shows the dependency of the audio-to-MIDI model on the downstream converter. The eight M21 pipelines occupy a narrow, uniformly poor interval from 96.1 to 96.9, a span of less than one OMR-NED point. By comparison, the MS and M2ST pipelines span more than five OMR-NED points across audio-to-MIDI models. A weak converter like M21 therefore largely masks differences among the upstream transcribers under OMR-NED. MUSTER's notation-oriented subscores give the same converter ordering on the same predicted and reference sheet music, so this ranking is not an OMR-NED artifact; full scores are in \supptab{tab:appendix-muster-raw-scores}{A.4}.

\subsubsection{Playback similarity}
\label{subsubsec:playback-system-analysis}

\begin{figure*}[t]
\centering
\includegraphics[width=0.8\textwidth]{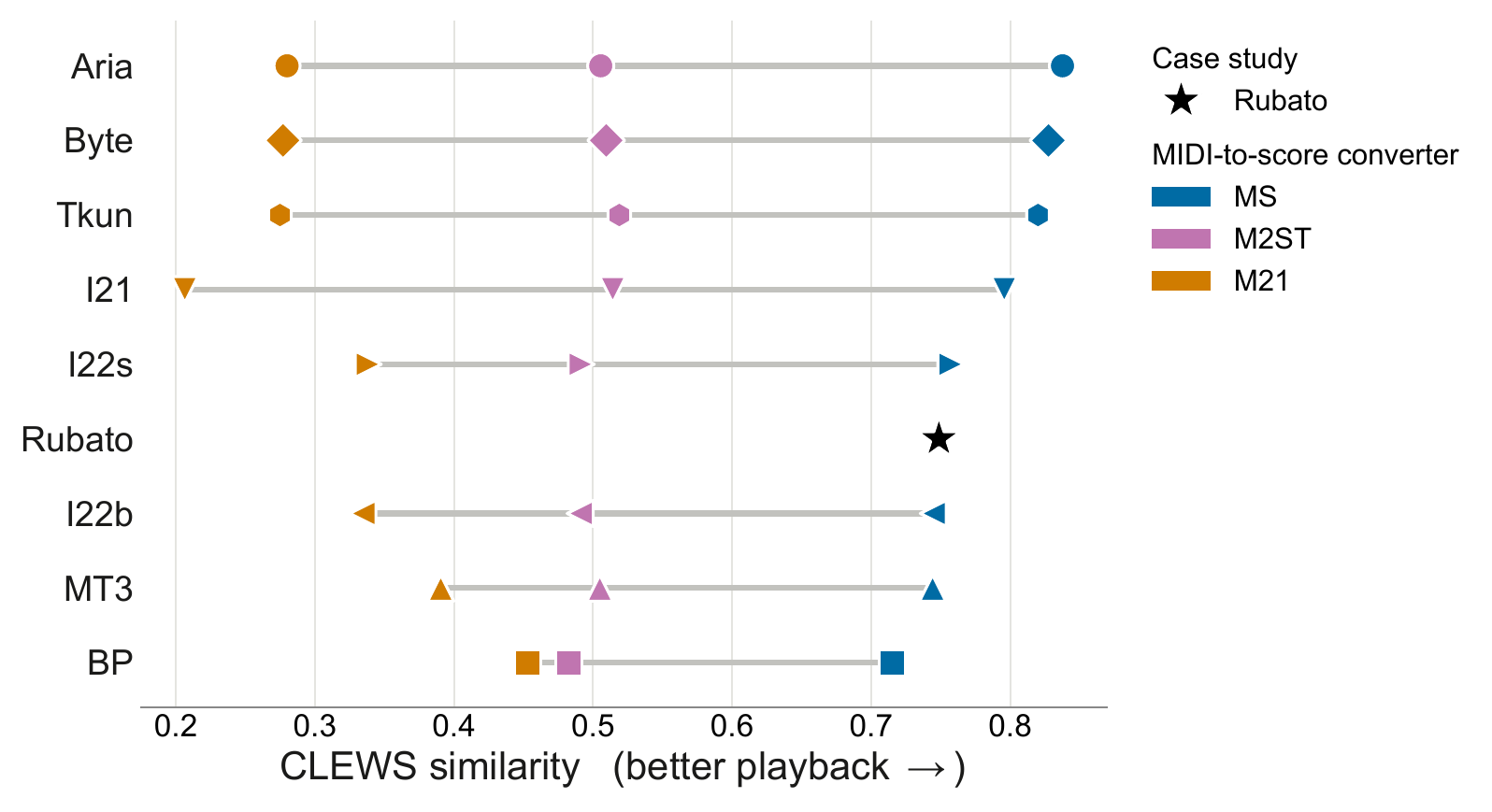}
\caption{Playback similarity across audio-to-score pipelines. Rows and marker shapes identify the audio-to-MIDI model, while marker colors identify the MIDI-to-score converter. The gray segment connects the three converter results for each transcriber. CLEWS similarity is higher-is-better, so stronger playback similarity appears to the right. Rows are ordered by the rightmost point in each row, strongest at the top; Rubato, shown separately as a case study with a black star, therefore falls into its overall rank of sixth.}
\label{fig:clews-system-analysis}
\end{figure*}

Figure~\ref{fig:clews-system-analysis} shows a different converter ordering for playback similarity. MS gives the highest CLEWS similarity for all eight audio-to-MIDI models, M2ST is intermediate, and M21 is weakest (and leads to a very different ranking). Aria+MS is the strongest system under CLEWS ($0.84$), followed by Byte+MS and Tkun+MS. The remaining playback evaluators show the same broad converter-level pattern, with MS-based pipelines occupying most of the leading positions.

\subsubsection{Effect of MIDI-to-score converter choice}
Taken together, the two figures show that the MIDI-to-score converter has two roles. First, it determines which property of the final output is favored: M2ST is strongest for notation similarity, whereas MS is strongest for playback similarity. Second, it controls how much of the variation among audio-to-MIDI models survives into the complete system. The near-collapse of all M21 pipelines into a uniformly weak OMR-NED range is the clearest example: differences among upstream transcribers produce little observable system-level separation when the downstream converter does not preserve them.

\subsection{Case study: Rubato}
\label{subsec:rubato-case-study}

Rubato~\citep{tamer2026rubatotranscribingpianomusic} differs from the 24 modular pipelines because it predicts time-aligned score notation directly from performance audio, with local context and without a separate MIDI-to-score converter. We therefore include it as an end-to-end case study. Rubato was released after our human listening study, so it is evaluated only with the available automatic metrics.

Rubato reaches the lowest OMR-NED of any system we evaluate, $72.30$ (Figure~\ref{fig:omr-system-analysis}), ahead of every modular pipeline including those built on M2ST, the strongest notation-side converter. We read this narrowly: predicting a \emph{time-aligned} score from audio, with a mechanism that keeps the prediction local to the audio it comes from, can produce more reference-like notation than a modular audio-to-MIDI plus MIDI-to-score pipeline.

Rubato also performs strongly in playback similarity (Figure~\ref{fig:clews-system-analysis}). It ranks sixth under CLEWS, and all five systems ranked above it use MS, which in turn are among the weaker systems on notational similarity to the reference. The remaining playback metrics show a similar pattern. The complete scores are reported in \supptab{tab:rubato-case-study-results}{A.6}.

In Figure~\ref{fig:omr-versus-playback-main}, Rubato sits outside the region traced by the MIDI-reliant systems: no modular pipeline beats it on both dimensions at once, while it beats 19 of the 24 on both, including every M2ST and M21 pipeline and three of the eight MS pipelines. The notation--playback tension we document among the modular systems is therefore a property of that design, not an inherent tradeoff: a system that predicts time-aligned notation can move outward on both axes at once.

\subsection{Validating playback metrics against human preference}
\label{subsec:human-validation}

We evaluate how closely the automatic playback metrics agree with human perception. We collect reference-anchored pairwise listening judgments (\S\ref{subsec:human-abx}), aggregate them into system-level Bradley--Terry scores (\S\ref{subsec:human-score-aggregation}), and measure rank correlation between these scores and each automatic playback evaluator. We first evaluate the DSP-based feature-alignment metrics (\S\ref{subsec:dsp-feature-alignment-validation}) and then the learned and model-based evaluators (\S\ref{subsec:model-based-playback-evaluation}). This analysis was used to select a representative variant within each evaluator family --- the feature for DTW, the feature and $(\nu,\lambda)$ setting for TWED, and the aggregation variant for CLEWS and CLaMP~3 --- and those selected variants are the rows of Table~\ref{tab:playback-evaluator-selection} and the columns of \supptab{tab:appendix-selected-view-raw-scores}{A.2}; it also determines which automatic metric to use in the main playback figures. Because the listening task evaluates rendered audio, it is used to validate playback-side metrics (not the notation-side OMR-NED metric).

\subsubsection{Human preference data collection}
\label{subsec:human-abx}

We collect human preference data using a reference-anchored two-alternative forced-choice listening task (sometimes called `ABX'). Each trial presents a labeled reference clip $X$, an excerpt of the original recording, and two candidate clips, $A$ and $B$, rendered from two different audio-to-score transcription pipelines and time-aligned to the same region of the piece. Playback starts at an assigned position within the recording and is not truncated: participants control playback and may replay any clip as often as they wish. Participants are asked to choose whether candidate $A$ or candidate $B$ better matches the reference (they are not shown any scores). We use the same rendering pipeline as in \S\ref{subsubsec:audio_rendering}, so human listeners and automatic playback-side metrics evaluate the same rendered audio.

The trial pool is defined by 24 evaluated pipelines, giving $\binom{24}{2}=276$ unordered system pairs, and 230 reference recordings. Each human judgment samples one reference recording and one system pair. Candidate order is randomized. For each trial, the \emph{side mapping} records which underlying pipeline is assigned to label $A$ and which is assigned to label $B$. The \emph{selected candidate} is the side, $A$ or $B$, chosen by the participant. Combining the selected candidate with the side mapping converts each response into one pairwise win/loss between two transcription pipelines.

Each participant completes 30 trials. Based on the simulation-based power analysis in Supplementary File, Appendix~A.6 and Table~A.8, our recruiting targeted 100 participants. The completed study contains 106 participants and 3,180 valid forced-choice judgments. We record confidence and playback counts as diagnostic metadata, but the main ranking analysis uses only the selected candidate and side mapping. Interface details, participant-background statistics, and diagnostic metadata are reported in \suppsec{app:human-evaluation-details}{A.7}.

\subsubsection{Human score aggregation}
\label{subsec:human-score-aggregation}

Different participants may make different choices for the same reference excerpt and system pair, especially when two rendered candidates are perceptually close. We therefore aggregate response-level pairwise judgments into one numerical human preference score for each transcription pipeline.

We use the Bradley--Terry model~\citep{bradley1952rank} as the aggregation method. Let $\theta_i$ denote the latent human preference score for pipeline $i$. For pipelines $i$ and $j$, the probability that pipeline $i$ is preferred over pipeline $j$ is modeled as
\begin{equation}
\label{eq:bradley-terry}
P(i \succ j)
=
\frac{\exp(\theta_i)}
{\exp(\theta_i)+\exp(\theta_j)}.
\end{equation}

To estimate $\theta$, we first convert each ABX response into a winner--loser pair using the selected candidate and the randomized side mapping. We then estimate $\theta$ by maximizing the Bradley--Terry log-likelihood over all observed pairwise wins,
\begin{equation}
\label{eq:bt-log-likelihood}
\mathcal{L}(\theta)
=
\sum_{(i,j)\in \mathcal{W}}
\log
\frac{\exp(\theta_i)}
{\exp(\theta_i)+\exp(\theta_j)},
\end{equation}
where $\mathcal{W}$ is the multiset of observed winner--loser pairs ($i, j)$. We impose the standard identifiability constraint $\sum_i \theta_i = 0$. The resulting maximum-likelihood estimate gives one human Bradley--Terry score per pipeline; larger $\theta_i$ indicates a stronger human preference. The raw Bradley--Terry scores are shown in \supptab{tab:appendix-selected-view-raw-scores}{A.2}.

\subsubsection{DSP-based feature-alignment metric validation}
\label{subsec:dsp-feature-alignment-validation}

We use Spearman correlation \citep{spearman1904proof} with Bradley--Terry scores as the primary selection criterion because it measures monotonic agreement between system-level rankings, and report Kendall's $\tau$ \citep{kendall1938new} as a secondary pairwise-agreement check. DTW and TWED costs are negated before correlation so higher is better.

DTW and TWED are deterministic DSP feature-alignment distances: they learn no parameters from our benchmark, but their behavior depends on the input feature representation. Applying each metric to each feature (one at a time), the best DTW setting is Chroma CENS (Spearman $\rho=0.891$, Kendall $\tau=0.761$), and the best TWED setting is MFCC with $(\nu,\lambda)=(0.001,2.0)$ ($\rho=0.924$, $\tau=0.790$). Table~\ref{tab:playback-evaluator-selection} summarizes the strongest evaluators; the full TWED sweep is in Supplementary File, Appendix~A.8 and the full per-feature DTW/TWED results in Appendix~A.9.

\begin{table*}[t]
\centering
\begin{adjustbox}{max width=\textwidth}
\begin{tabular}{llcc}
\toprule
Evaluator & Representation / input & Spearman $\rho$ $\uparrow$ & Kendall $\tau$ $\uparrow$ \\
\midrule
CLEWS & segment-mean audio embedding & \textbf{0.971} & \textbf{0.891} \\
Gemini 3.1 Pro & Audio ABX prompt & 0.970 & 0.870 \\
TWED & MFCC; selected cost negated & 0.924 & 0.790 \\
DTW & Chroma CENS; cost negated & 0.891 & 0.761 \\
CLaMP~3 audio & segment-median audio embedding & 0.884 & 0.710 \\
CLaMP~3 audio$\rightarrow$ABC & median ref-audio/pred-ABC cosine & 0.879 & 0.710 \\
CLaMP~3 ABC$\rightarrow$audio & mean ref-ABC/pred-audio cosine & 0.849 & 0.717 \\
\bottomrule
\end{tabular}
\end{adjustbox}

\caption{Automatic playback-side evaluators ranked by rank correlation with Bradley--Terry scores. For DTW and TWED we report the strongest feature (Chroma CENS and MFCC, respectively). Higher correlation is better.}
\label{tab:playback-evaluator-selection}
\end{table*}

\subsubsection{Model-based and embedding-based evaluation validation}
\label{subsec:model-based-playback-evaluation}

The model-based and embedding-based results in Table~\ref{tab:playback-evaluator-selection} compare the evaluators defined in \S\ref{subsubsec:model-based-metrics} against the same Bradley--Terry scores. CLEWS segment-mean similarity shows the strongest agreement with human preference (Spearman $\rho=0.971$, Kendall $\tau=0.891$), narrowly followed by the position-balanced Gemini 3.1 Pro judge ($\rho=0.970$, $\tau=0.870$). CLaMP~3 audio segment-median similarity also shows strong agreement ($\rho=0.884$, $\tau=0.710$), a little below DTW. Full model- and variant-level results are reported in \suppsec{app:supplementary-correlations}{A.9}.

Although CLaMP~3 ABC does not compare rendered audio, we also evaluate its agreement with human playback preference to examine the effect of CLaMP~3's shared multimodal representation. Its global cosine score correlates more strongly with human preference (Spearman $\rho=0.824$, Kendall $\tau=0.659$) than with OMR-NED ($\rho=0.681$, $\tau=0.478$). This suggests that the shared embedding emphasizes modality-invariant musical content even when its input is symbolic. We therefore interpret CLaMP~3 ABC as a measure of broad musical correspondence, not as an explicit notation or playback metric.

\subsection{Cost analysis}
\label{subsec:evaluation-cost}

Agreement with human preference is only one criterion for selecting an evaluator; cost, reproducibility, and coverage also matter. Local notation, DSP, and embedding metrics score all reference--candidate pairs, whereas the human and AudioLM studies use sampled pairwise comparisons. Table \ref{tab:evaluation-cost-comparison} therefore reports each method's cost per 1,000 comparison units.

\begin{table*}[t]
\centering
\begin{adjustbox}{max width=\textwidth}
\begin{tabular}{llrrr}
\toprule
Evaluator & Unit & Num.\ units & Cost estimate & Normalized cost \\
\midrule
OMR-NED & symbolic-score pair & 5,520 & \$1.10 & \$0.200 / 1,000 units \\
\midrule
DTW & audio-pair score & 5,520 & \$3.37 & \$0.610 / 1,000 units \\
TWED--MFCC & audio-pair score & 5,520 & \$9.74 & \$1.760 / 1,000 units \\
CLaMP~3 audio & audio-pair score & 5,520 & \$15.45 & \$2.800 / 1,000 units \\
CLaMP~3 audio/ABC & audio--symbolic pair & 5,424 & \$3.99 & \$0.740 / 1,000 units \\
CLaMP~3 ABC & symbolic-score pair & 5,261 & \$0.95 & \$0.180 / 1,000 units \\
CLEWS & audio-pair score & 5,520 & \$2.06 & \$0.370 / 1,000 units \\
Gemini 3.1 Pro & ABX request (both orders) & 3,600 & \$18.36 & \$5.100 / 1,000 units \\
Human ABX & judgment & 3,180 & \$1,060.00 & \$333.333 / 1,000 units \\
\bottomrule
\end{tabular}
\end{adjustbox}

\caption{Evaluation cost across notation, feature-alignment, learned-embedding, AudioLM, and human evaluation. Dollar amounts for local metrics are cloud-equivalent estimates; AudioLM costs are paid-API estimates; human ABX is participant compensation. Normalized values are useful within this benchmark but are not exact like-for-like comparisons across evaluator families.}
\label{tab:evaluation-cost-comparison}
\end{table*}

\begin{figure*}[t]
\centering
\includegraphics[width=0.8\textwidth]{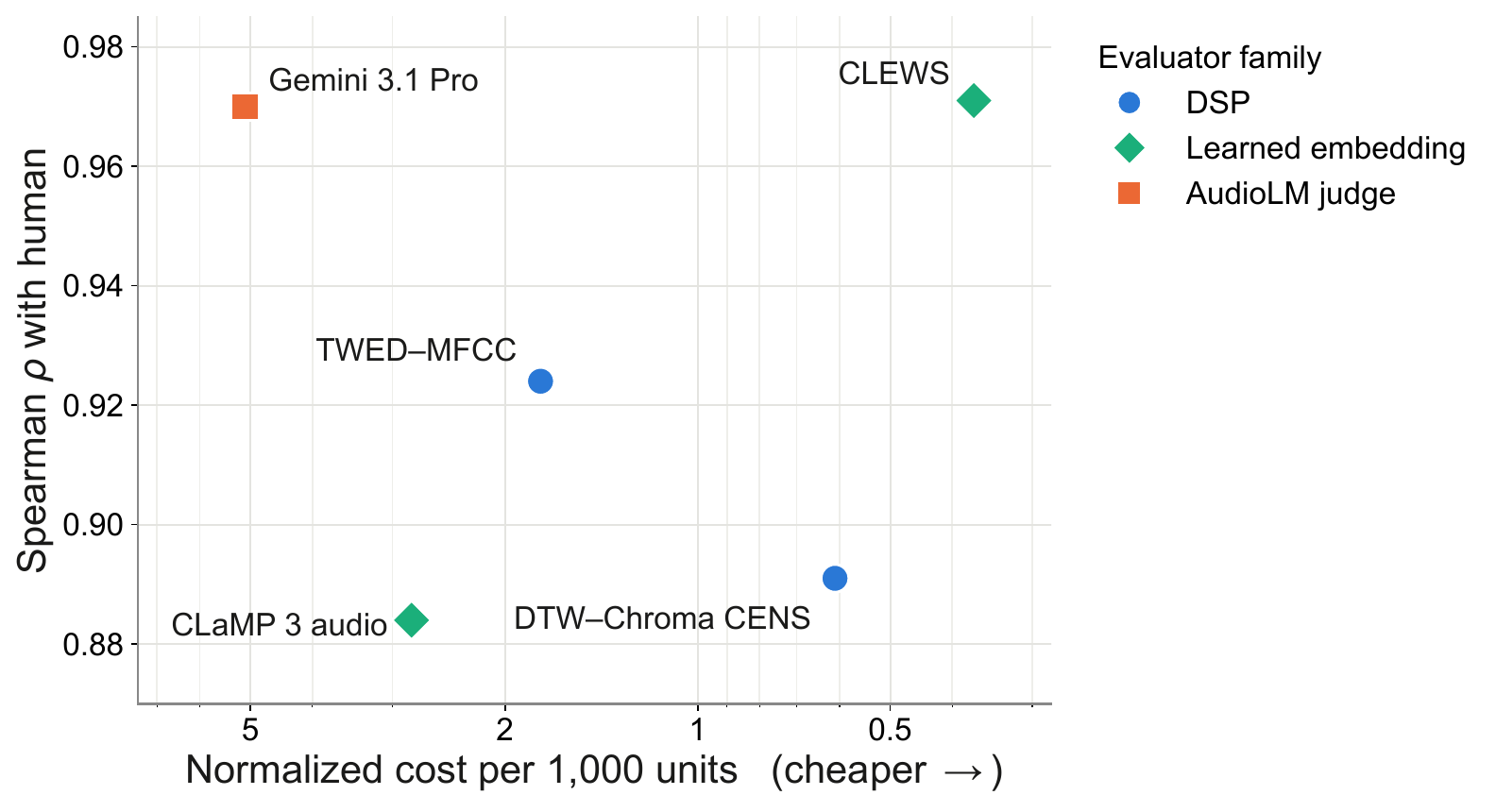}
\caption{Cost--human agreement tradeoff for five automatic playback similarity evaluators. Cost is normalized per 1,000 evaluation units using Table~\ref{tab:evaluation-cost-comparison}; the horizontal axis is reversed so that cost decreases to the right, and the vertical axis reports Spearman correlation with human-derived Bradley--Terry scores using Table~\ref{tab:playback-evaluator-selection}. Better evaluators are therefore up and to the right, as in the other figures.}
\label{fig:playback-cost-vs-human-correlation}
\end{figure*}

Figure~\ref{fig:playback-cost-vs-human-correlation} makes the practical tradeoff explicit. CLEWS provides the strongest cost--agreement balance, combining the highest human correlation with the lowest normalized cost among the five playback evaluators. DTW remains attractive when a DSP, model-independent screen is preferred. TWED increases agreement relative to DTW at higher compute cost, while Gemini approaches CLEWS's agreement at the highest automatic-evaluation cost shown. CLaMP~3 audio provides another learned representation view, but does not improve the cost--agreement frontier here.

\section{Discussion}
\label{sec:discussion}

For AMT research, we recommend reporting and comparing measures of both notation similarity and playback similarity.  For reasons discussed in \S\ref{subsec:related-amt-evaluation}, we recommend OMR-NED for the former, at this writing. On the playback side, CLEWS (segment-mean) is the strongest choice among the automatic playback similarity evaluators we considered, in terms of both correlation to human preferences and cost (Figure~\ref{fig:playback-cost-vs-human-correlation}).  If a stable, model-free option is required, TWED--MFCC or DTW--Chroma CENS should be considered, depending on budget.  We do not recommend AudioLM judges at this writing; while they can provide ABX-format preference comparisons that approximate human judgments almost as well as CLEWS, they are $\sim$13$\times$ as expensive and depend on model version, prompt, and parsing choices (as well as proprietary and unstable model weights, in many cases).
Our methodology in \S\ref{subsec:human-validation} can be applied to assess new proposals for automatic playback similarity evaluation.

\section{Limitations}

This study focuses on Western classical piano recordings, evaluated over the first three minutes of each recording; conclusions may not transfer to other instruments, ensembles, voice, non-classical genres, or longer structures. Playback-side results depend on the rendering pipeline: although all systems are rendered consistently, the renderer, soundfont, and normalization could affect both automatic metrics and listeners. We also do not validate notation quality against human score-reading judgments; OMR-NED measures symbol similarity but may not reflect the effort to read or edit generated sheet music. Finally, our human-validated comparison targets pipelined systems; we take a first step toward end-to-end audio-to-score models with the Rubato case study (\S\ref{subsec:rubato-case-study}), but Rubato postdates the listening study and so lacks human judgments, leaving its perceived playback quality an open question. Future work might add human visual evaluation of readability and test the stability of metric selection across held-out pieces, participants, and systems.

\section{Conclusion}
\label{sec:conclusion}

We presented a dual evaluation of audio-to-score transcription that separates written-score similarity from rendered-playback similarity. These two views rank systems differently because they measure distinct properties of the output.
In general, we recommend that music transcription evaluations include methods to assess both the extent to which the generated sheet music resembles the reference and the extent to which it preserves the musical content of the original recording when rendered, and future transcription systems intended for diverse use should consider test-time controls that allow a user to trade off between these two different goals.

\ifpreprint

\begin{fundinginfo}[Funding information]
Participant compensation, reported in Table~\ref{tab:evaluation-cost-comparison}, was covered by the last author's discretionary research funding.
\end{fundinginfo}

\fi

\section*{Competing Interests}
The authors have no competing interests to declare.

\section*{Authors' Contributions}
\ifpreprint
PW designed and implemented the evaluation pipeline, ran the transcription, rendering and metric computations, designed and deployed the human listening study, analysed the results, and led the writing of the manuscript. GY contributed to the OMR-NED experiments, advised on OMR-NED and the notation-side evaluation, and provided ongoing methodological guidance throughout the project. NCT advised on the Rubato case study and on the CLaMP~3 and CLEWS embedding evaluators. VE advised on the manuscript and its figures and on the analysis of the results. NAS conceived the project, supervised it, and served as principal investigator on the IRB protocol.  All authors contributed to the framing and writing of the paper.
\else
Author contributions are withheld for double-blind review and are recorded in the online submission system.
\fi

\section*{Use of Generative AI}
Generative AI tools were used to check grammar and wording, and to assist with writing parts of the evaluation and analysis code, which the authors reviewed and tested. They were not used to generate research ideas, design the evaluation, produce or interpret any result, or write the scientific content. Audio language models were separately \emph{evaluated} as candidate playback-similarity metrics, as described in \S\ref{subsubsec:model-based-metrics}.

\section*{Ethics and Consent}
This study included a human listening experiment (\S\ref{subsec:human-abx}) with 106 participants aged 18 or over, conducted in accordance with the Declaration of Helsinki.
\ifpreprint
The protocol, `Human Evaluation of Automatic Music Transcription Systems', was reviewed by the University of Washington Human Subjects Division, which on 12 March 2026 determined it to be human subjects research qualifying for exempt status under Category~2 (IRB~ID STUDY00025192).
\else
The protocol was reviewed by the authors' institutional Human Subjects Division, which determined it to be human subjects research qualifying for exempt status under Category~2; the institution, IRB reference number and determination letter are withheld here for double-blind review and have been supplied to the editors.
\fi
Each participant gave informed consent through an electronic form stating the purpose, procedure, duration, risks, benefits, compensation and voluntary nature of the study, and could stop at any time without penalty; documentation of written consent was waived for this anonymous, minimal-risk online experiment. No directly identifying information was collected, and compensation was a gift certificate requiring none.

\section*{Reproducibility}
\ifpreprint
The evaluation code, the de-identified human ABX response data and the per-system scores are released at \url{https://github.com/pingw220/AMT-Dual-Eval} under the MIT License, with the released data tables under CC~BY~4.0. The release covers the rendered-audio (DTW/TWED) dimension and the human listening study; it excludes reference and rendered audio, MusicXML and MIDI outputs, and third-party model outputs, which remain subject to their own licences. Participant identifiers are replaced with surrogate ids and the mapping is not released; demographics appear only as aggregate counts.
\else
The evaluation code, the de-identified human ABX responses and the per-system scores are publicly released in a GitHub repository under the MIT License, with the data tables under CC~BY~4.0. Its URL is withheld for double-blind review and has been supplied to the editors. Reference and rendered audio, MusicXML and MIDI outputs, and third-party model outputs are excluded, remaining subject to their own licences. Participant identifiers are replaced with surrogate ids and demographics appear only as aggregate counts.
\fi

\ifpreprint
\begin{authoraff}[AUTHOR AFFILIATIONS]
\end{authoraff}
\fi
\clearpage

\bibliographystyle{apaTISMIR}
\bibliography{references}

\ifpreprint
  \clearpage
    %TC:ignore
    \input{appendix.tex}
    %TC:endignore
\fi

\nolinenumbers

\label{endPage_Label}

\end{document}

%% file: appendix.tex
% Appendix body. Deliberately has no preamble and does not \input the main
% file: main.tex must be the single top-level document, or arXiv cannot tell
% which file to compile.
\twocolumn[%
  {\raggedright\sffamily\bfseries\fontsize{17bp}{21bp}\selectfont
   Supplementary File: Appendix\par}%
  \vspace{10pt}%
  {\raggedright\fontsize{11bp}{15bp}\selectfont
   Supplementary Material for \emph{A Dual Evaluation for Music Transcription}\par}%
  \vspace{22pt}%
]

\appendix

\renewcommand{\tablename}{Table}
\renewcommand{\thetable}{A.\arabic{table}}

\begingroup
\setlength{\abovecaptionskip}{8pt}
\setlength{\belowcaptionskip}{8pt}
\setlength{\textfloatsep}{14pt plus 2pt minus 2pt}
\setlength{\floatsep}{14pt plus 2pt minus 2pt}
\setlength{\intextsep}{12pt plus 2pt minus 2pt}
\renewcommand{\arraystretch}{1.06}

\section{OMR-NED Definition Details}
\label{app:omr-ned-details}

This section gives the full definition of OMR-NED summarised in \mainsec{subsec:notation_similarity}{2.1}.

OMR-NED defines edit operations over music-notation symbols. These symbols include note and rest properties, such as pitch, accidentals, note heads, beams, flags, dots, ties, articulations, and ornaments, as well as non-note score objects such as clefs, key signatures, time signatures, dynamics, slurs, lyrics, directions, chord symbols, and other notation elements. The metric is therefore intended to measure differences in the sheet music that would require manual correction by a user of an OMR (or, in our case, AMT) system.

Let $\hat{Y}$ denote a transcription and $Y$ denote the reference sheet music. Consider a transformation from $\hat{Y}$ into $Y$ through insertions and deletions of notation symbols; let $I(\hat{Y},Y)$ be the number of insertions and $D(\hat{Y},Y)$ the number of deletions in the transformation that minimizes their sum. (OMR-NED does not count substitutions as a separate operation; a changed symbol is represented as one deletion and one insertion.)  The per-score OMR-NED is defined as
\begin{equation}
\mathrm{OMR\text{-}NED}(\hat{Y},Y)
=
\frac{I(\hat{Y},Y) + D(\hat{Y},Y)}
{|\hat{Y}| + |Y|},
\end{equation}
where $|\cdot|$ denotes the number of notation-symbols in the argument; this normalization makes OMR-NED length-aware by scaling the edit distance by the total amount of notation present in both documents. Lower OMR-NED indicates stronger notational similarity, and a value of zero indicates an exact match.

Following \citet{martinezsevilla2025smb}, our implementation reports a micro-averaged OMR-NED over all evaluated files:
\begin{equation}
\mathrm{Overall\ OMR\text{-}NED}
=
\frac{
\sum_k I(\hat{Y}_k,Y_k) + \sum_k D(\hat{Y}_k,Y_k)
}{
\sum_k |\hat{Y}_k|+ \sum_k |Y_k|
},
\end{equation}
where $k$ indexes evaluated recordings and sheet-music pairs. For readability, all reported OMR-NED values, in both tables and figures, are scaled by $100$.

\section{Supplementary System Raw Scores and Rankings}
\label{app:supplementary-system-tables}

This section reports complete 24-system raw scores and ranks. Table~\ref{tab:appendix-selected-view-raw-scores} reports the raw values for the selected notation, rendered-audio, preference-based, and learned-embedding evaluators, while Table~\ref{tab:appendix-selected-view-ranks} reports the corresponding rank positions. Table~\ref{tab:appendix-muster-correlations} compares each MUSTER subscore with OMR-NED and human playback preference; Tables~\ref{tab:appendix-muster-raw-scores} and~\ref{tab:appendix-muster-ranks} provide the underlying full-system results.

\begin{table}[t]
\centering
\small
\setlength{\tabcolsep}{4pt}
\begin{tabular}{lcc}
\toprule
MUSTER score & OMR-NED & Human playback \\
\midrule
MeanER   & 0.383  & 0.792 \\
PitchER  & $-0.345$ & 0.405 \\
MissRate & $-0.219$ & 0.481 \\
ExtraRate& 0.044  & 0.657 \\
OnsetER  & 0.504  & 0.838 \\
OffsetER & 0.782  & 0.761 \\
VoiceER  & 0.944  & 0.454 \\
HandER   & 0.812  & 0.160 \\
ScaleErr & 0.807  & 0.570 \\
\bottomrule
\end{tabular}

\caption{Spearman rank correlations of MUSTER subscores with OMR-NED and human
Bradley--Terry playback preference over the 24 modular systems. Metric
directions are oriented consistently before correlation, so a larger positive
value indicates more similar system rankings. MUSTER's structural subscores
are most notation-like, whereas its official MeanER and several
note-oriented subscores are more playback-like.}
\label{tab:appendix-muster-correlations}
\end{table}

\begin{table*}[tbp]
\centering
\footnotesize
\setlength{\tabcolsep}{3pt}
\begin{tabular}{lcccccccc}
\toprule
System & OMR-NED $\downarrow$ & DTW $\downarrow$ & TWED $\downarrow$ & Human $\uparrow$ & Gemini $\uparrow$ & CLaMP~3 audio $\uparrow$ & CLEWS $\uparrow$ & CLaMP~3 ABC $\uparrow$ \\
\midrule
Aria+MS & 91.12 & \textbf{0.07973} & 0.42597 & \textbf{1.8907} & \textbf{2.4788} & 0.94295 & \textbf{0.83759} & 0.57698 \\
Byte+MS & 91.08 & 0.08245 & 0.42747 & 1.5421 & 1.7055 & \textbf{0.94357} & 0.82751 & 0.54236 \\
Tkun+MS & 92.61 & 0.09000 & 0.43853 & 1.4451 & 1.9714 & 0.93913 & 0.82000 & \textbf{0.58417} \\
I21+MS & 95.62 & 0.09705 & \textbf{0.41390} & 1.2087 & 1.6280 & 0.93987 & 0.79570 & 0.41130 \\
I22s+MS & 90.29 & 0.10894 & 0.66980 & 0.6326 & 0.6080 & 0.92467 & 0.75739 & 0.57271 \\
MT3+MS & 90.01 & 0.11255 & 0.66232 & 0.4156 & 0.6053 & 0.92803 & 0.74418 & 0.57880 \\
Aria+M2ST & 87.15 & 0.09936 & 0.80503 & 0.3969 & 0.1748 & 0.90913 & 0.50553 & 0.55672 \\
I22b+MS & 90.18 & 0.11928 & 0.67648 & 0.3362 & 0.4916 & 0.91856 & 0.74467 & 0.57868 \\
Tkun+M2ST & \textbf{86.61} & 0.09837 & 0.79908 & 0.3250 & 0.2966 & 0.90989 & 0.51888 & 0.57511 \\
Byte+M2ST & 87.91 & 0.10128 & 0.82107 & 0.2817 & 0.2129 & 0.91093 & 0.50951 & 0.53643 \\
BP+MS & 91.67 & 0.09787 & 0.55742 & 0.1635 & 0.5717 & 0.93376 & 0.71498 & 0.49885 \\
MT3+M2ST & 89.37 & 0.11892 & 0.78900 & 0.0508 & 0.0345 & 0.91147 & 0.50489 & 0.49061 \\
I21+M2ST & 88.33 & 0.10576 & 0.78970 & 0.0028 & -0.1251 & 0.91172 & 0.51415 & 0.53774 \\
I22s+M2ST & 89.25 & 0.12321 & 0.83003 & -0.2044 & -0.2662 & 0.90933 & 0.49155 & 0.48314 \\
BP+M21 & 96.65 & 0.12085 & 0.71550 & -0.3065 & -0.5816 & 0.90784 & 0.45297 & 0.20880 \\
I22b+M2ST & 89.12 & 0.12186 & 0.82004 & -0.4171 & -0.2199 & 0.90959 & 0.49074 & 0.48107 \\
BP+M2ST & 91.83 & 0.10139 & 0.81752 & -0.5361 & -0.2556 & 0.91495 & 0.48274 & 0.43430 \\
I22s+M21 & 96.27 & 0.17904 & 2.54533 & -0.6349 & -1.1535 & 0.85619 & 0.33837 & 0.29025 \\
Aria+M21 & 96.35 & 0.19105 & 4.11370 & -0.7131 & -1.1686 & 0.83579 & 0.27996 & 0.29833 \\
MT3+M21 & 96.07 & 0.15901 & 1.41916 & -0.7879 & -1.0516 & 0.88494 & 0.39048 & 0.27035 \\
I22b+M21 & 96.26 & 0.18865 & 2.77154 & -0.8999 & -1.0498 & 0.85170 & 0.33449 & 0.27909 \\
Byte+M21 & 96.33 & 0.19549 & 3.66545 & -1.0096 & -1.3184 & 0.84242 & 0.27706 & 0.28428 \\
Tkun+M21 & 96.60 & 0.20767 & 5.37409 & -1.3852 & -1.2622 & 0.82417 & 0.27488 & 0.30908 \\
I21+M21 & 96.90 & 0.27218 & 7.20457 & -1.7970 & -2.3265 & 0.79850 & 0.20644 & 0.23558 \\
\bottomrule
\end{tabular}

\caption{Full selected-view raw scores for all 24 systems. Lower OMR-NED, DTW, and TWED values indicate better scores; higher human Bradley--Terry, position-balanced Gemini 3.1 Pro Bradley--Terry, CLaMP~3 audio, CLEWS, and CLaMP~3 ABC scores indicate stronger preference or similarity. Best values within each metric column are bolded. CLaMP~3 audio uses segment-median cosine similarity, CLEWS uses segment-mean similarity, and CLaMP~3 ABC uses median global cosine similarity.}
\label{tab:appendix-selected-view-raw-scores}
\end{table*}

The raw scores in Table~\ref{tab:appendix-selected-view-raw-scores} show that the best system depends strongly on the evaluation view: Tkun+M2ST is best under OMR-NED, Aria+MS is best under DTW, human ABX, position-balanced Gemini 3.1 Pro, and CLEWS, I21+MS is best under TWED, Byte+MS is best under CLaMP~3 audio, and Tkun+MS is best under CLaMP~3 ABC.

\begin{table*}[tbp]
\centering
\footnotesize
\setlength{\tabcolsep}{3pt}
\begin{tabular}{lcccccccc}
\toprule
System & OMR-NED & DTW & TWED & Human ABX & Gemini 3.1 Pro & CLaMP~3 audio & CLEWS & CLaMP~3 ABC \\
\midrule
Aria+MS & 12 & \textbf{1} & 2 & \textbf{1} & \textbf{1} & 2 & \textbf{1} & 4 \\
Byte+MS & 11 & 2 & 3 & 2 & 3 & \textbf{1} & 2 & 8 \\
Tkun+MS & 15 & 3 & 4 & 3 & 2 & 4 & 3 & \textbf{1} \\
I21+MS & 16 & 4 & \textbf{1} & 4 & 4 & 3 & 4 & 16 \\
I22s+MS & 10 & 11 & 7 & 5 & 5 & 7 & 5 & 6 \\
MT3+MS & 8 & 12 & 6 & 6 & 6 & 6 & 7 & 2 \\
Aria+M2ST & 2 & 7 & 13 & 7 & 11 & 16 & 12 & 7 \\
I22b+MS & 9 & 14 & 8 & 8 & 8 & 8 & 6 & 3 \\
Tkun+M2ST & \textbf{1} & 6 & 12 & 9 & 9 & 13 & 9 & 5 \\
Byte+M2ST & 3 & 8 & 16 & 10 & 10 & 12 & 11 & 10 \\
BP+MS & 13 & 5 & 5 & 11 & 7 & 5 & 8 & 11 \\
MT3+M2ST & 7 & 13 & 10 & 12 & 12 & 11 & 13 & 12 \\
I21+M2ST & 4 & 10 & 11 & 13 & 13 & 10 & 10 & 9 \\
I22s+M2ST & 6 & 17 & 17 & 14 & 16 & 15 & 14 & 13 \\
BP+M21 & 23 & 15 & 9 & 15 & 17 & 17 & 17 & 24 \\
I22b+M2ST & 5 & 16 & 15 & 16 & 14 & 14 & 15 & 14 \\
BP+M2ST & 14 & 9 & 14 & 17 & 15 & 9 & 16 & 15 \\
I22s+M21 & 19 & 19 & 19 & 18 & 20 & 19 & 19 & 19 \\
Aria+M21 & 21 & 21 & 22 & 19 & 21 & 22 & 21 & 18 \\
MT3+M21 & 17 & 18 & 18 & 20 & 19 & 18 & 18 & 22 \\
I22b+M21 & 18 & 20 & 20 & 21 & 18 & 20 & 20 & 21 \\
Byte+M21 & 20 & 22 & 21 & 22 & 23 & 21 & 22 & 20 \\
Tkun+M21 & 22 & 23 & 23 & 23 & 22 & 23 & 23 & 17 \\
I21+M21 & 24 & 24 & 24 & 24 & 24 & 24 & 24 & 23 \\
\bottomrule
\end{tabular}

\caption{Full selected-view rank table for all 24 systems. Lower rank is better for every column. The CLaMP~3 audio, CLEWS, and CLaMP~3 ABC columns use the same representative score variants as Table~\ref{tab:appendix-selected-view-raw-scores}. Best rank values are bolded.}
\label{tab:appendix-selected-view-ranks}
\end{table*}

Table~\ref{tab:appendix-selected-view-ranks} reports the corresponding rank positions and makes the converter-level split clear: M2ST systems rank highest under OMR-NED, while MS systems rank highest under playback-side, preference-based, and learned audio-embedding views.

\begin{table*}[tbp]
\centering
\footnotesize
\setlength{\tabcolsep}{2.5pt}
\begin{adjustbox}{max width=\textwidth}
\begin{tabular}{lccccccccc}
\toprule
System & MeanER & PitchER & MissRate & ExtraRate & OnsetER & OffsetER & VoiceER & HandER & ScaleErr \\
\midrule
Aria+MS & 24.82 & 2.70 & 8.35 & 25.02 & 36.09 & 51.92 & 73.51 & 58.63 & 1.64 \\
Byte+MS & 27.13 & 3.79 & 10.62 & 29.14 & 36.77 & 55.35 & 73.77 & 55.93 & 1.69 \\
Tkun+MS & \textbf{22.65} & \textbf{1.68} & \textbf{6.32} & \textbf{21.48} & 40.44 & 43.33 & 71.99 & 58.03 & 1.40 \\
I21+MS & 32.96 & 5.49 & 11.49 & 34.20 & 50.76 & 62.86 & 75.71 & 57.88 & 1.98 \\
I22s+MS & 25.67 & 2.69 & 18.17 & 25.32 & 34.29 & 47.89 & 71.37 & 56.28 & 1.58 \\
MT3+MS & 22.78 & 7.05 & 12.16 & 28.62 & \textbf{28.05} & \textbf{38.00} & 67.56 & 55.17 & \textbf{1.37} \\
I22b+MS & 25.45 & 2.57 & 17.87 & 25.31 & 34.35 & 47.16 & 69.92 & 55.31 & 1.55 \\
BP+MS & 37.93 & 7.97 & 34.42 & 37.49 & 47.62 & 62.15 & 75.21 & 57.75 & 1.79 \\
Aria+M2ST & 41.13 & 20.13 & 36.00 & 47.06 & 54.30 & 48.18 & 58.51 & 47.30 & 1.48 \\
Tkun+M2ST & 40.90 & 19.66 & 36.70 & 46.21 & 53.88 & 48.06 & \textbf{55.25} & \textbf{44.61} & 1.47 \\
Byte+M2ST & 42.94 & 22.62 & 35.80 & 48.25 & 57.08 & 50.98 & 61.32 & 49.97 & 1.54 \\
MT3+M2ST & 42.92 & 24.49 & 37.24 & 46.70 & 54.98 & 51.20 & 66.81 & 56.20 & 1.52 \\
I21+M2ST & 46.04 & 25.17 & 37.22 & 52.73 & 59.69 & 55.40 & 61.93 & 49.17 & 1.61 \\
I22s+M2ST & 42.76 & 20.99 & 42.25 & 43.55 & 54.38 & 52.61 & 63.30 & 50.97 & 1.58 \\
I22b+M2ST & 42.99 & 21.15 & 42.38 & 44.11 & 55.01 & 52.29 & 65.40 & 53.55 & 1.56 \\
BP+M2ST & 48.08 & 25.54 & 45.41 & 45.81 & 62.32 & 61.31 & 63.65 & 50.46 & 1.73 \\
BP+M21 & 49.65 & 13.80 & 40.65 & 43.24 & 70.51 & 80.06 & 87.51 & 70.78 & 2.05 \\
I22s+M21 & 48.98 & 11.77 & 36.23 & 46.01 & 71.82 & 79.06 & 78.40 & 56.52 & 1.97 \\
Aria+M21 & 47.04 & 11.16 & 28.65 & 47.33 & 65.29 & 82.76 & 80.36 & 60.86 & 2.04 \\
MT3+M21 & 49.82 & 16.47 & 34.96 & 50.55 & 73.46 & 73.67 & 75.11 & 51.77 & 1.79 \\
I22b+M21 & 49.58 & 11.89 & 37.11 & 47.16 & 72.36 & 79.38 & 77.41 & 55.48 & 1.95 \\
Byte+M21 & 47.96 & 13.09 & 29.54 & 48.54 & 66.53 & 82.08 & 80.54 & 57.22 & 2.05 \\
Tkun+M21 & 40.18 & 6.54 & 20.48 & 35.02 & 59.04 & 79.80 & 77.34 & 59.10 & 1.66 \\
I21+M21 & 47.52 & 12.26 & 27.27 & 49.98 & 60.96 & 87.15 & 81.44 & 66.01 & 2.30 \\
\bottomrule
\end{tabular}
\end{adjustbox}

\caption{Full MUSTER system-level error rates for the 24 modular systems. All values are percentages and lower is better. MeanER is the unweighted mean of PitchER, MissRate, ExtraRate, OnsetER, and OffsetER.}
\label{tab:appendix-muster-raw-scores}
\end{table*}

\begin{table*}[tbp]
\centering
\footnotesize
\setlength{\tabcolsep}{3pt}
\begin{tabular}{lccccccccc}
\toprule
System & MeanER & PitchER & MissRate & ExtraRate & OnsetER & OffsetER & VoiceER & HandER & ScaleErr \\
\midrule
Aria+MS & 3 & 4 & 2 & 2 & 4 & 9 & 13 & 20 & 12 \\
Byte+MS & 6 & 5 & 3 & 6 & 5 & 12 & 14 & 12 & 14 \\
Tkun+MS & \textbf{1} & \textbf{1} & \textbf{1} & \textbf{1} & 6 & 2 & 12 & 19 & 2 \\
I21+MS & 7 & 6 & 4 & 7 & 8 & 16 & 17 & 18 & 20 \\
I22s+MS & 5 & 3 & 7 & 4 & 2 & 4 & 11 & 14 & 10 \\
MT3+MS & 2 & 8 & 5 & 5 & \textbf{1} & \textbf{1} & 9 & 9 & \textbf{1} \\
I22b+MS & 4 & 2 & 6 & 3 & 3 & 3 & 10 & 10 & 7 \\
BP+MS & 8 & 9 & 12 & 9 & 7 & 15 & 16 & 17 & 17 \\
Aria+M2ST & 11 & 18 & 15 & 17 & 10 & 6 & 2 & 2 & 4 \\
Tkun+M2ST & 10 & 17 & 17 & 15 & 9 & 5 & \textbf{1} & \textbf{1} & 3 \\
Byte+M2ST & 14 & 21 & 14 & 20 & 14 & 7 & 3 & 4 & 6 \\
MT3+M2ST & 13 & 22 & 20 & 16 & 12 & 8 & 8 & 13 & 5 \\
I21+M2ST & 16 & 23 & 19 & 24 & 16 & 13 & 4 & 3 & 11 \\
I22s+M2ST & 12 & 19 & 22 & 11 & 11 & 11 & 5 & 6 & 9 \\
I22b+M2ST & 15 & 20 & 23 & 12 & 13 & 10 & 7 & 8 & 8 \\
BP+M2ST & 20 & 24 & 24 & 13 & 18 & 14 & 6 & 5 & 15 \\
BP+M21 & 23 & 15 & 21 & 10 & 21 & 21 & 24 & 24 & 22 \\
I22s+M21 & 21 & 11 & 16 & 14 & 22 & 18 & 20 & 15 & 19 \\
Aria+M21 & 17 & 10 & 10 & 19 & 19 & 23 & 21 & 22 & 21 \\
MT3+M21 & 24 & 16 & 13 & 23 & 24 & 17 & 15 & 7 & 16 \\
I22b+M21 & 22 & 12 & 18 & 18 & 23 & 19 & 19 & 11 & 18 \\
Byte+M21 & 19 & 14 & 11 & 21 & 20 & 22 & 22 & 16 & 23 \\
Tkun+M21 & 9 & 7 & 8 & 8 & 15 & 20 & 18 & 21 & 13 \\
I21+M21 & 18 & 13 & 9 & 22 & 17 & 24 & 23 & 23 & 24 \\
\bottomrule
\end{tabular}

\caption{Full MUSTER ranks for the 24 modular systems. Lower is better. Rubato is excluded from this ranking and reported separately in Table~\ref{tab:rubato-case-study-results}.}
\label{tab:appendix-muster-ranks}
\end{table*}

\begin{table}[t]
\centering
\small
\setlength{\tabcolsep}{4pt}
\begin{tabular}{llcc}
\toprule
Metric & Side & Score & Rank/25 \\
\midrule
OMR-NED $\downarrow$ & notation & 72.30 & 1 \\
CLaMP~3 ABC $\uparrow$ & notation & 0.658 & 1 \\
MUSTER MeanER $\downarrow$ & notation & 12.28 & 1 \\
MUSTER PitchER $\downarrow$ & notation & 3.30 & 5 \\
MUSTER MissRate $\downarrow$ & notation & 8.52 & 3 \\
MUSTER ExtraRate $\downarrow$ & notation & 24.88 & 2 \\
MUSTER OnsetER $\downarrow$ & notation & 8.25 & 1 \\
MUSTER OffsetER $\downarrow$ & notation & 16.46 & 1 \\
MUSTER VoiceER $\downarrow$ & notation & 45.76 & 1 \\
MUSTER HandER $\downarrow$ & notation & 38.56 & 1 \\
MUSTER ScaleErr $\downarrow$ & notation & 1.16 & 1 \\
\midrule
Gemini 3.1 Pro $\uparrow$ & playback & 1.777 & 4 \\
TWED--MFCC $\downarrow$ & playback & 0.639 & 6 \\
CLEWS $\uparrow$ & playback & 0.749 & 6 \\
CLaMP~3 audio $\uparrow$ & playback & 0.928 & 7 \\
DTW--Chroma CENS $\downarrow$ & playback & 0.123 & 17 \\
\bottomrule
\end{tabular}

\caption{Rubato case-study results. Ranks are computed among the 24 modular systems plus Rubato for automatic metrics. Rubato has no human ABX score because it was not included in the original human pairwise-comparison pool. Lower values are better for OMR-NED, MUSTER, DTW, and TWED; higher values are better for CLaMP~3, CLEWS, and Gemini.}
\label{tab:rubato-case-study-results}
\end{table}

\section{Extended Related Work on Transcription Evaluation}
\label{app:extended-related-work}

This section expands the related-work summaries in \mainsec{subsec:related-amt-evaluation}{2.1.1} and \mainsec{subsec:related-audio-faithfulness}{2.2.5}.

\subsection{Notation-side evaluation}
\label{app:extended-related-notation}

\textbf{\citet{cogliati2017metric}} propose a music-notation edit-distance metric that compares predicted and reference MusicXML scores across multiple notation dimensions, including notes, durations, rests, barlines, signatures, groupings, and staff assignment. This provides a detailed analysis of notation errors. In addition, they train a linear-regression model to map these notation-error counts to human ratings of pitch notation, rhythm notation, and note positioning. This learned mapping is estimated on a relatively small human-rated dataset, and the metric does not enforce disjoint penalties: one underlying structural mistake, such as an incorrect meter or quantization error, may lead to multiple counted notation errors.

\textbf{MV2H}~\citep{mcleod2018mv2h,mcleod2019nonaligned} evaluates five dimensions of complete transcription: multi-pitch detection, voice separation, meter, note value, and harmony. Its disjoint-penalty design aims to penalize each underlying transcription error only once. For example, voice separation and note-value accuracy are evaluated only on appropriate matched notes, so a missing note is not repeatedly penalized as both a pitch error and a downstream voice or note-value error. MV2H is designed mainly to evaluate the correctness of musical structure, such as detected notes, voices, metrical structure, note values, and harmonic labels. It places less emphasis on the readability of the final sheet music and on visual or typesetting properties such as beaming, stem directions, grouping, rests, and staff placement. Moreover, its official MusicXML evaluation workflow first converts the reference and predicted scores to MIDI, so the comparison is performed over a MIDI-derived event and timing representation rather than directly over the written notation objects in the MusicXML scores. Its overall score is also an unweighted average of distinct evaluation dimensions.

\textbf{MUSTER}~\citep{nakamura2018towards,hiramatsu2021joint} evaluates transcriptions through separate error rates for pitch, missing and extra notes, onsets and offsets, voice assignment, voice continuity, rhythmic scale, and hand or staff assignment. Its official MeanER is the unweighted mean of five note-oriented measures: PitchER, MissRate, ExtraRate, OnsetER, and OffsetER. The remaining subscores, including VoiceER, HandER, and ScaleErr, describe different aspects of score organization and are not included in MeanER.

We applied MUSTER to the same predicted and reference sheet music as a comparison with OMR-NED and playback-side evaluations. Its subscores do not support a consistent interpretation (Table~\ref{tab:appendix-muster-correlations}). The structure-oriented VoiceER, HandER, and ScaleErr rankings correlate strongly with OMR-NED ($\rho=0.94$, $0.81$, and $0.81$, respectively), suggesting that they primarily reflect notation-side structure. In contrast, MeanER and the note-oriented OnsetER correlate much more strongly with human playback preference as described in \mainsec{subsec:human-validation}{4.4} ($\rho=0.79$ and $0.84$) than with OMR-NED ($\rho=0.38$ and $0.50$). PitchER, MissRate, and ExtraRate show the same tendency. MUSTER is a notation-evaluation framework in terms of its inputs and intended task, but its subscores emphasize different objectives and can make its system ranking look either notation-like or playback-like.

This inconsistency makes MUSTER useful for reporting specific error types, but unsuitable as this paper's primary notation similarity metric: selecting a different subscore changes the conclusion, while MeanER omits several score-organization properties central to the visual notation task. We therefore use OMR-NED for the main notation comparison. Full MUSTER scores and ranks are reported in Tables~\ref{tab:appendix-muster-raw-scores} and~\ref{tab:appendix-muster-ranks}.

\subsection{Playback-side evaluation}
\label{app:extended-related-playback}

\textbf{\citet{ycart2020perceptual}} investigate the perceptual validity of automatic piano transcription metrics through a pairwise listening test, in which listeners choose which of two rendered MIDI transcriptions sounds more similar to a rendered ground-truth MIDI reference. They further train a model from these human judgments using features computed from symbolic target and transcription output. This work is closely related to our playback-side evaluation, but differs from our setting in two important ways. First, their learned metric is trained from human evaluations of short excerpts from MAPS, a piano-transcription dataset with aligned audio and MIDI annotations~\citep{emiya2010maps}, produced by four MIDI-level AMT systems. Second, their listening task compares synthesized MIDI transcriptions against a synthesized MIDI reference. In contrast, we compare audio rendered from final MusicXML scores directly against the original recorded performances. Our evaluation therefore measures whether generated sheet music preserves the musical content heard in the original recording, rather than whether it sounds similar to a synthesized symbolic reference.

\textbf{\citet{simonetta2022perceptual}} further study perceptual evaluation of resynthesized AMT outputs and show that standard objective AMT metrics computed from MIDI note events may correlate weakly with listener judgments of preserved musical interpretation. These prior studies motivate evaluating not only note-event correctness, but also the perceptual faithfulness of rendered transcription outputs.

\section{Audio Feature Details}
\label{app:audio-feature-details}

This section lists the frame-level audio features used for the DTW and TWED evaluation summarised in \mainsec{subsubsec:audio-features}{2.2.2}.

\begin{table*}[t]
\centering
\begin{adjustbox}{max width=\textwidth}
\begin{tabular}{llll}
\toprule
Feature & Citation & Feature group & Main musical property \\
\midrule
Chroma & \citep{muller2011chroma} & Pitch/harmony & Pitch-class content \\
HPCP & \citep{gomez2006tonal} & Pitch/harmony & Harmonic pitch-class profile \\
Chroma CENS & \citep{muller2011chroma} & Pitch/harmony & Smoothed pitch-class structure \\
Tonnetz & \citep{harte2006detecting} & Pitch/harmony & Tonal and harmonic relations \\
CQT & \citep{brown1991calculation} & Pitch/harmony & Log-frequency pitch structure \\
STFT semitone & \citep{allen1977short,mcfee2015librosa} & Pitch/harmony & Semitone-level spectral pitch content \\
Pitch salience & \citep{salamon2012melody,bogdanov2013essentia} & Pitch/harmony & Prominent pitch activity \\
\midrule
Mel spectrogram & \citep{slaney1998auditory,mcfee2015librosa} & Spectral & Perceptual frequency-band energy \\
MFCC & \citep{davis1980comparison} & Spectral & Spectral envelope and cepstral structure \\
PCEN-Mel & \citep{wang2017trainable} & Spectral & Compressed and normalized Mel energy \\
Log-STFT & \citep{allen1977short} & Spectral & Log-scaled short-time spectrum \\
Spectral flux & \citep{bello2005tutorial} & Spectral/onset & Frame-to-frame spectral change \\
\midrule
Onset strength & \citep{bello2005tutorial,ellis2007beat} & Rhythm/onset & Note-onset activity \\
Tempogram & \citep{grosche2010cyclic} & Rhythm/onset & Rhythmic periodicity and tempo structure \\
\bottomrule
\end{tabular}
\end{adjustbox}

\caption{Audio features used for DTW and TWED evaluation. The experiment uses 14 audio features. Each feature is cited to either a standard MIR formulation, the underlying signal representation, or the analysis toolkit used in our implementation.}
\label{tab:audio-features}
\end{table*}

\section{DTW and TWED Details}
\label{app:dtw-twed-details}

This appendix gives the full definitions of the DTW and TWED distances used for playback-side evaluation. In the main paper, we summarize these methods because both are standard time-series alignment algorithms; here, we specify the exact scoring direction and normalization used in our implementation.

\subsection{Dynamic time warping}
\label{app:dtw-details}

Dynamic time warping~(DTW) measures the minimum alignment cost between two temporal feature sequences. Let
\[
X = \langle x_1,\ldots,x_n\rangle
\]
be the reference feature sequence, and let
\[
Y = \langle y_1,\ldots,y_m\rangle
\]
be the candidate feature sequence. DTW computes an alignment path
\[
P = \langle (i_1,j_1), (i_2,j_2), \ldots, (i_K,j_K)\rangle
\]
that minimizes the accumulated local cost
\[
\sum_{k=1}^{K} d(x_{i_k}, y_{j_k}),
\]
subject to the boundary conditions $(i_1,j_1)=(1,1)$ and $(i_K,j_K)=(n,m)$, and the step-size condition
\[
(i_k-i_{k-1}, j_k-j_{k-1}) \in \{(1,0),(0,1),(1,1)\}
\]
for all $k=2,\ldots,K$.

In our implementation, DTW is computed over frame-level audio features extracted from the original reference recording and from the audio rendered from each candidate score. Given feature vectors $x_i$ and $y_j$, the local cost is cosine distance:
\begin{equation}
d(x_i,y_j)
=
1 -
\frac{\langle x_i, y_j\rangle}
{\|x_i\|_2 \|y_j\|_2}.
\end{equation}

The reported DTW cost is normalized by the length of the selected warping path:
\begin{equation}
\mathrm{DTWCost}(X,Y)
=
\frac{1}{K}
\sum_{k=1}^{K}
d(x_{i_k}, y_{j_k}),
\end{equation}
where $K$ is the DTW path length. Lower DTW cost indicates more similar audio feature trajectories. For correlation analysis with higher-is-better human and model-based scores, we negate the distance:
\begin{equation}
\mathrm{DTWSim}(X,Y)
=
-\mathrm{DTWCost}(X,Y).
\end{equation}

\subsection{Time warp edit distance}
\label{app:twed-details}

Time warp edit distance~(TWED) compares time-stamped feature sequences using temporal warping together with edit-style insertion/deletion costs. Let
\[
X = \langle x_1,\ldots,x_n\rangle,
\qquad
Y = \langle y_1,\ldots,y_m\rangle
\]
be two feature sequences with timestamps $t^X_i$ and $t^Y_j$. TWED computes a dynamic program over match, deletion-from-$X$, and deletion-from-$Y$ operations. For $i,j>0$, the recurrence is:
\begin{align}
\mathrm{match}
&=
D(i-1,j-1)
+
\lVert x_i-y_j\rVert_2
+
\nu |t^X_i-t^Y_j|,
\\
\mathrm{deleteX}
&=
D(i-1,j)
+
\lVert x_i-x_{i-1}\rVert_2
+
\lambda
+
\nu(t^X_i-t^X_{i-1}),
\\
\mathrm{deleteY}
&=
D(i,j-1)
+
\lVert y_j-y_{j-1}\rVert_2
+
\lambda
+
\nu(t^Y_j-t^Y_{j-1}),
\\
D(i,j)
&=
\min\{\mathrm{match},\mathrm{deleteX},\mathrm{deleteY}\}.
\end{align}

The parameter $\lambda$ controls the edit/deletion penalty, while $\nu$ controls temporal stiffness. Larger $\lambda$ penalizes insertions and deletions more strongly, and larger $\nu$ makes the distance more sensitive to timestamp differences.

We evaluate TWED under four parameter settings:
\[
(\nu,\lambda)
\in
\{(0.001,1.0), (0.01,1.0), (0.001,2.0), (0.01,2.0)\}.
\]
In all settings, TWED is computed over frame-level audio features using frame-index timestamps, frame-wise $\ell_2$ normalization, and Euclidean local distance. Lower TWED cost indicates stronger rendered-audio similarity. As with DTW, we negate TWED costs when computing rank correlations with higher-is-better human and model-based scores:
\begin{equation}
\mathrm{TWEDSim}(X,Y)
=
-\mathrm{TWEDCost}(X,Y)
=
-D(n,m).
\end{equation}

\section{Simulation-based Power Analysis}
\label{app:power-analysis}

We conducted a simulation-based power analysis to choose the sample sizes for human and audio-language-model pairwise evaluation. The analysis simulates the same reference-anchored pairwise comparison setting used in the main experiments: 24 transcription systems, 30 trials per participant or automated-judge block, randomized A/B order, and aggregation of pairwise outcomes into system-level rankings.

The simulation defines power as ranking-recovery probability rather than as a null-hypothesis rejection rate. A simulated study is counted as successful if the recovered ranking satisfies a composite stability criterion. The baseline criterion requires ranking stability, top-5 stability, and agreement across aggregation methods. We report the binary, medium-strictness, baseline-success setting as the primary operating point because it matches the forced-choice human listening interface and is more conservative than a ternary design with ties.

Table~\ref{tab:appendix-power-analysis} reports the resulting power curve for the operating point used to choose the human and audio-language-model evaluation sizes. The columns with 80 and 300 simulations are two separate Monte Carlo estimates of the same design point, not a lower and upper bound. The 300-simulation run has lower Monte Carlo error and is the value cited in the main paper. In this setting, $N=60$ gives a 300-simulation power estimate of 0.870, while $N=80$ gives 0.920 and meets the 0.90 target.

\begin{table*}[t]
\centering
\small
\begin{tabular}{cccc}
\toprule
$N$ & Total judgments & Power, 80 simulations & Power, 300 simulations \\
\midrule
40  & 1,200 & 0.763 & 0.763 \\
60  & 1,800 & 0.825 & 0.870 \\
80  & 2,400 & 0.925 & 0.920 \\
100 & 3,000 & 0.938 & 0.947 \\
\bottomrule
\end{tabular}

\caption{Power curve for the conservative binary, medium-strictness, baseline-success configuration. Each simulated participant or judge block contributes 30 trials. The 80-simulation and 300-simulation columns are separate Monte Carlo estimates of the same design point; the 300-simulation estimate is the reported value.}
\label{tab:appendix-power-analysis}
\end{table*}

The minimum tested sample size meeting the 0.90 target under the conservative setting is $N=80$. We therefore treat $N=80$ as the powered minimum for human evaluation and target approximately $N=100$ to provide additional robustness to incomplete participation or response-quality variation. The final human study slightly exceeds this target, with 106 participants and 3,180 valid judgments. For audio-language-model evaluation, we use $N=60$, or approximately 1,800 pairwise requests. This setting reaches the 0.80-power regime under the conservative binary criterion and is sufficient for automated-judge comparison, while human judgments remain the primary perceptual reference.

\noindent\textbf{Interpreting the confidence intervals.}
Power is the mean of binary simulation outcomes: each simulated study either passes or fails the composite recovery criterion. The standard deviation of these 0/1 outcomes is not the uncertainty in the power estimate. The uncertainty that should be reported is the Monte Carlo standard error or, preferably, the bootstrap confidence interval for the power estimate. For example, the 300-simulation run estimates power of 0.870 at $N=60$, with bootstrap 95\% confidence interval $[0.830,0.907]$; at $N=80$, it estimates power of 0.920, with bootstrap 95\% confidence interval $[0.890,0.950]$.

\section{Supplementary Human Evaluation Details}
\label{app:human-evaluation-details}

This section provides additional details about the human ABX evaluation described in \mainsec{subsec:human-abx}{4.4.1}. The main paper reports the task design and study size, while this appendix summarizes interface details, participant metadata, diagnostic response fields, and the assignment schedule.

\subsection{Human ABX interface and instructions}
\label{app:human-abx-interface}

The human listening interface labels the clips only as $X$ (the reference), $A$, and $B$; model and converter identities are hidden from participants. Candidate presentation order is randomized independently for each trial. The instructions ask participants to focus on pitch, onset/timing, and duration, and to ignore timbre, instrument sound quality, audio realism, loudness, reverb, random noise artifacts, and overall prettiness. This instruction is important because all candidate clips are synthesized renderings of generated scores, and our goal is to measure whether the transcription preserves the musical content of the original performance rather than whether the rendering sounds realistic.

The deployed task uses forced $A/B$ choices with no tie option. If two clips are close, participants are instructed to choose the slightly closer option and report low confidence. We record the selected side, the randomized side mapping, participant confidence, response time, and playback counts for the reference and candidate clips. Confidence, response time, and playback counts are used only for diagnostics in this paper; the main Bradley--Terry aggregation uses the selected side and side mapping.

\subsection{Trial-assignment schedule}
\label{app:human-trial-assignment}

The human ABX assignment schedule is generated before deployment. The full trial pool contains 24 evaluated transcription pipelines, giving $\binom{24}{2}=276$ unordered system pairs, and 230 reference recordings. Each participant template contains 30 trials. Each trial specifies a reference recording, a pair of systems, a randomized assignment of the two systems to candidate labels $A$ and $B$, and an excerpt start percentage. Playback starts at the assigned position and is not truncated: participants listen for as long as they like and may replay any clip.

The assignment schedule is over-provisioned relative to the target number of completed participants. This makes the design robust to incomplete recruitment: the first completed participants still form a valid prefix of the planned assignment schedule. In the full schedule, each unordered system pair appears approximately equally often, so no system pair dominates the collected judgments. The schedule also balances four excerpt start positions, corresponding to $0\%$, $25\%$, $50\%$, and $75\%$ of the evaluated portion of the recording, which is capped at 3 minutes as in the automatic evaluation.

For each assigned trial, the start percentage is converted into start times for the reference and both candidate clips. The assignment file stores the aligned offsets as \texttt{ref\_start\_sec}, \texttt{candA\_start\_sec}, and \texttt{candB\_start\_sec}. These offsets are used by the frontend to start playback from the assigned region. This prevents all judgments from concentrating only on the beginning of each excerpt and helps distribute listening comparisons across different musical regions.

Table~\ref{tab:appendix-human-study-summary} summarizes the final human ABX study size, response format, and diagnostic metadata recorded by the interface.

\begin{table*}[t]
\centering
\small
\begin{tabular}{lc}
\toprule
Quantity & Value \\
\midrule
Completed participants & 106 \\
Trials per participant & 30 \\
Valid forced-choice judgments & 3,180 \\
Evaluated systems & 24 \\
Unordered system pairs & 276 \\
Reference recordings & 230 \\
Candidate labels & $A$ and $B$ \\
Reference label & $X$ \\
Response format & Forced $A/B$ choice \\
Tie option & No \\
Primary aggregation method & Bradley--Terry \\
Diagnostic metadata & Confidence, response time, playback counts \\
\bottomrule
\end{tabular}

\caption{Summary of the human ABX evaluation.}
\label{tab:appendix-human-study-summary}
\end{table*}

\subsection{Participant background}
\label{app:participant-background}

In addition to response-level ABX choices, we collect participants' self-reported music background, including proficiency level and years of music training. Table~\ref{tab:appendix-participant-background-table} summarizes self-reported proficiency level and years of music training. The participant pool contains substantial musical experience: 13 participants self-report as music majors, 75 as amateur musicians, and 18 as having no musical experience. Table~\ref{tab:appendix-proficiency-training-crosstab} further breaks down years of training among participants who self-report as amateur musicians or music majors.

\begin{table}[t]
\centering
\small
\begin{tabular}{llc}
\toprule
Category & Group & Count \\
\midrule
\multirow{4}{*}{Proficiency level}
& No experience & 18 \\
& Amateur & 75 \\
& Music major & 13 \\
& Missing / prefer not to answer & 0 \\
\midrule
\multirow{6}{*}{Years of music training}
& 0 years & 16 \\
& 1--3 years & 16 \\
& 4--7 years & 25 \\
& 8--10 years & 22 \\
& More than 10 years & 27 \\
& Missing / prefer not to answer & 0 \\
\bottomrule
\end{tabular}

\caption{Participant music-background summary ($N=106$ completed participants): self-reported proficiency level and years of music training.}
\label{tab:appendix-participant-background-table}
\end{table}

\begin{table*}[t]
\centering
\small
\begin{tabular}{lccc}
\toprule
Years of music training & Music major & Amateur & Combined \\
\midrule
0 years & 0 & 3 & 3 \\
1--3 years & 1 & 12 & 13 \\
4--7 years & 2 & 21 & 23 \\
8--10 years & 2 & 20 & 22 \\
More than 10 years & 8 & 19 & 27 \\
\midrule
Total & 13 & 75 & 88 \\
\bottomrule
\end{tabular}

\caption{Years of music training among participants who self-report as amateur musicians or music majors. The combined column shows the music-experienced subset of the participant pool.}
\label{tab:appendix-proficiency-training-crosstab}
\end{table*}

\section{Supplementary TWED Parameter Analysis}
\label{app:twed-parameter-selection}

We evaluate four TWED parameter configurations to determine whether the temporal-stiffness parameter $\nu$ and edit-penalty parameter $\lambda$ affect agreement with human Bradley--Terry scores. For each parameter configuration, we select the strongest feature under the primary Spearman rank-correlation criterion. This section supplements the TWED method description in \mainsec{subsubsec:feature-alignment-metrics}{2.2.3} and the evaluator-selection discussion in \mainsec{subsec:human-validation}{4.4}.

Table~\ref{tab:twed-parameter-sweep} reports the best TWED feature under each tested parameter configuration.

\begin{table}[t]
\centering
\small
\begin{tabular}{cccc}
\toprule
$\nu$ & $\lambda$ & Spearman $\rho$ $\uparrow$ & Kendall $\tau$ $\uparrow$ \\
\midrule
0.001  & 1.0 & 0.908 & 0.761 \\
0.01   & 1.0 & 0.908 & 0.761 \\
\textbf{0.001} & \textbf{2.0} & \textbf{0.924} & \textbf{0.790} \\
0.01   & 2.0 & \textbf{0.924} & \textbf{0.790} \\
\bottomrule
\end{tabular}

\caption{Best TWED feature under each tested parameter configuration, measured against human Bradley--Terry scores. MFCC is the strongest feature for every tested configuration. Higher correlation values are better.}
\label{tab:twed-parameter-sweep}
\end{table}

The two MFCC configurations with $\lambda=2.0$ tie under the primary human Bradley--Terry comparison. We select $\nu=0.001$ and $\lambda=2.0$ for the main paper because it is more robust across all 14 evaluated feature representations: its mean Spearman correlation across features is $0.889$, compared with $0.875$ for $\nu=0.01$ and $\lambda=2.0$.

\section{Supplementary Metric Correlation Results}
\label{app:supplementary-correlations}

This section provides the full metric-correlation results supporting the evaluator-selection analysis in \mainsec{subsec:human-validation}{4.4}. Table~\ref{tab:appendix-dtw-twed-feature-correlations} compares DTW and TWED using the same feature rows. Table~\ref{tab:appendix-audiolm-correlations} reports the AudioLM judge correlations. Table~\ref{tab:appendix-learned-embedding-correlations} combines the CLaMP~3 audio, CLEWS, and CLaMP~3 ABC correlation results in a single learned-embedding table.

\begin{table*}[tbp]
\centering
\small
\setlength{\tabcolsep}{5pt}
\begin{tabular}{lcccc}
\toprule
Feature & DTW Spearman $\rho$ & DTW Kendall $\tau$ & TWED Spearman $\rho$ & TWED Kendall $\tau$ \\
\midrule
Chroma & 0.872 & 0.739 & 0.891 & 0.761 \\
HPCP & 0.872 & 0.739 & 0.891 & 0.761 \\
Chroma CENS & \textbf{0.891} & \textbf{0.761} & 0.879 & 0.739 \\
Tonnetz & 0.864 & 0.717 & 0.857 & 0.703 \\
CQT & 0.726 & 0.587 & 0.870 & 0.725 \\
STFT semitone & 0.777 & 0.645 & 0.907 & 0.783 \\
Pitch salience & 0.710 & 0.572 & 0.903 & 0.768 \\
Mel spectrogram & 0.794 & 0.645 & 0.874 & 0.717 \\
MFCC & 0.743 & 0.536 & \textbf{0.924} & \textbf{0.790} \\
PCEN-Mel & 0.830 & 0.674 & 0.910 & 0.768 \\
Log-STFT & 0.796 & 0.645 & 0.904 & 0.775 \\
Spectral flux & 0.777 & 0.623 & 0.861 & 0.696 \\
Onset strength & 0.465 & 0.304 & 0.865 & 0.703 \\
Tempogram & 0.880 & 0.710 & 0.912 & 0.761 \\
\bottomrule
\end{tabular}

\caption{Full DTW and TWED feature-level correlations with human Bradley--Terry scores. DTW and TWED costs are negated before correlation so that higher values indicate better rendered-audio similarity. TWED uses $\nu=0.001$ and $\lambda=2.0$. Higher correlation values are better.}
\label{tab:appendix-dtw-twed-feature-correlations}
\end{table*}

Table~\ref{tab:appendix-dtw-twed-feature-correlations} shows that DTW and TWED favor different feature representations. DTW is strongest with Chroma CENS, while TWED is strongest with MFCC. DTW is also more sensitive to feature choice, whereas TWED remains highly correlated with human preference across most tested features.

\begin{table*}[t]
\centering
\small
\begin{tabular}{lcc}
\toprule
AudioLM judge & Spearman $\rho$ $\uparrow$ & Kendall $\tau$ $\uparrow$ \\
\midrule
Gemini 3.1 Pro preview (position-balanced) & \textbf{0.970} & \textbf{0.870} \\
Gemini 3 Pro preview & 0.964 & \textbf{0.870} \\
Gemini 3.1 Pro preview & 0.958 & 0.848 \\
Gemini 2.5 Pro & 0.825 & 0.681 \\
Gemini 2.5 Flash & 0.617 & 0.428 \\
Gemini 2.0 Flash-Lite & 0.507 & 0.377 \\
Gemini 3.1 Flash-Lite preview & 0.257 & 0.181 \\
Gemini 2.0 Flash & 0.035 & 0.022 \\
Gemini 3 Flash preview & -0.052 & -0.022 \\
Gemini 2.5 Flash-Lite & -0.127 & -0.080 \\
\bottomrule
\end{tabular}

\caption{Full AudioLM judge correlations with human Bradley--Terry scores. Higher values indicate stronger agreement with human preference.}
\label{tab:appendix-audiolm-correlations}
\end{table*}

Table~\ref{tab:appendix-audiolm-correlations} distinguishes the position-balanced Gemini 3.1 Pro run used in the main results from the single-order screening runs. Position balancing produces the strongest Spearman correlation, while the Gemini 3 Pro and Gemini 3.1 Pro screening runs are also strongly aligned with human Bradley--Terry scores. Smaller or lower-cost variants are substantially less aligned.

\begin{table*}[tbp]
\centering
\small
\begin{tabular}{llcc}
\toprule
Evaluator family & Score variant & Spearman $\rho$ $\uparrow$ & Kendall $\tau$ $\uparrow$ \\
\midrule
\multirow{4}{*}{CLaMP~3 audio}
& Segment median cosine & \textbf{0.884} & \textbf{0.710} \\
& Segment mean cosine & 0.872 & 0.688 \\
& Best-match mean cosine & 0.619 & 0.471 \\
& Global cosine & 0.247 & 0.145 \\
\midrule
\multirow{6}{*}{CLEWS}
& Segment mean similarity & \textbf{0.971} & \textbf{0.891} \\
& Segment median similarity & 0.967 & 0.884 \\
& Global cosine similarity & 0.950 & 0.848 \\
& Best-match mean similarity & 0.936 & 0.826 \\
& Mean similarity & 0.934 & 0.819 \\
& Mean distance & 0.930 & 0.812 \\
\midrule
\multirow{1}{*}{CLaMP~3 ABC}
& Global cosine similarity & \textbf{0.824} & \textbf{0.659} \\
\midrule
\multirow{2}{*}{CLaMP~3 audio$\rightarrow$ABC}
& Segment median similarity & \textbf{0.879} & \textbf{0.710} \\
& Segment mean similarity & 0.865 & 0.703 \\
\multirow{1}{*}{CLaMP~3 ABC$\rightarrow$audio}
& Segment mean similarity & \textbf{0.849} & \textbf{0.717} \\
\bottomrule
\end{tabular}

\caption{Combined learned-embedding correlations with human Bradley--Terry scores. CLaMP~3 audio and CLEWS compare reference audio with rendered candidate audio, while CLaMP~3 ABC compares score-derived ABC representations. Higher values indicate stronger agreement with human preference. Best values within each evaluator family are bolded.}
\label{tab:appendix-learned-embedding-correlations}
\end{table*}

Table~\ref{tab:appendix-learned-embedding-correlations} shows that CLEWS segment-mean similarity is the strongest learned-embedding evaluator in this benchmark. CLaMP~3 audio also aligns strongly with human preference when segment-level aggregation is used, whereas CLaMP~3 ABC is weaker because it compares score-derived symbolic representations rather than rendered audio. For CLaMP~3 ABC, conversion and embedding extraction succeeded on 5,261 of 5,520 expected score rows, giving an overall coverage of 95.3\%. We treat this success rate as a diagnostic coverage statistic rather than an evaluation score.

\section{Supplementary System Analysis Figures}
\label{app:supplementary-system-figures}

Figure~\ref{fig:appendix-system-ranges-rubato} extends the system analysis in
\mainsec{subsec:pipeline-component-effects}{4.2} from OMR-NED and CLEWS to all selected
automatic notation and playback evaluators, with Rubato included as a separate
end-to-end system. Rubato has no human Bradley--Terry score because it was
released after the listening study, so the human preference view covers only the
24 modular pipelines.

\begin{figure*}[tbp]
\centering
\includegraphics[width=\textwidth]{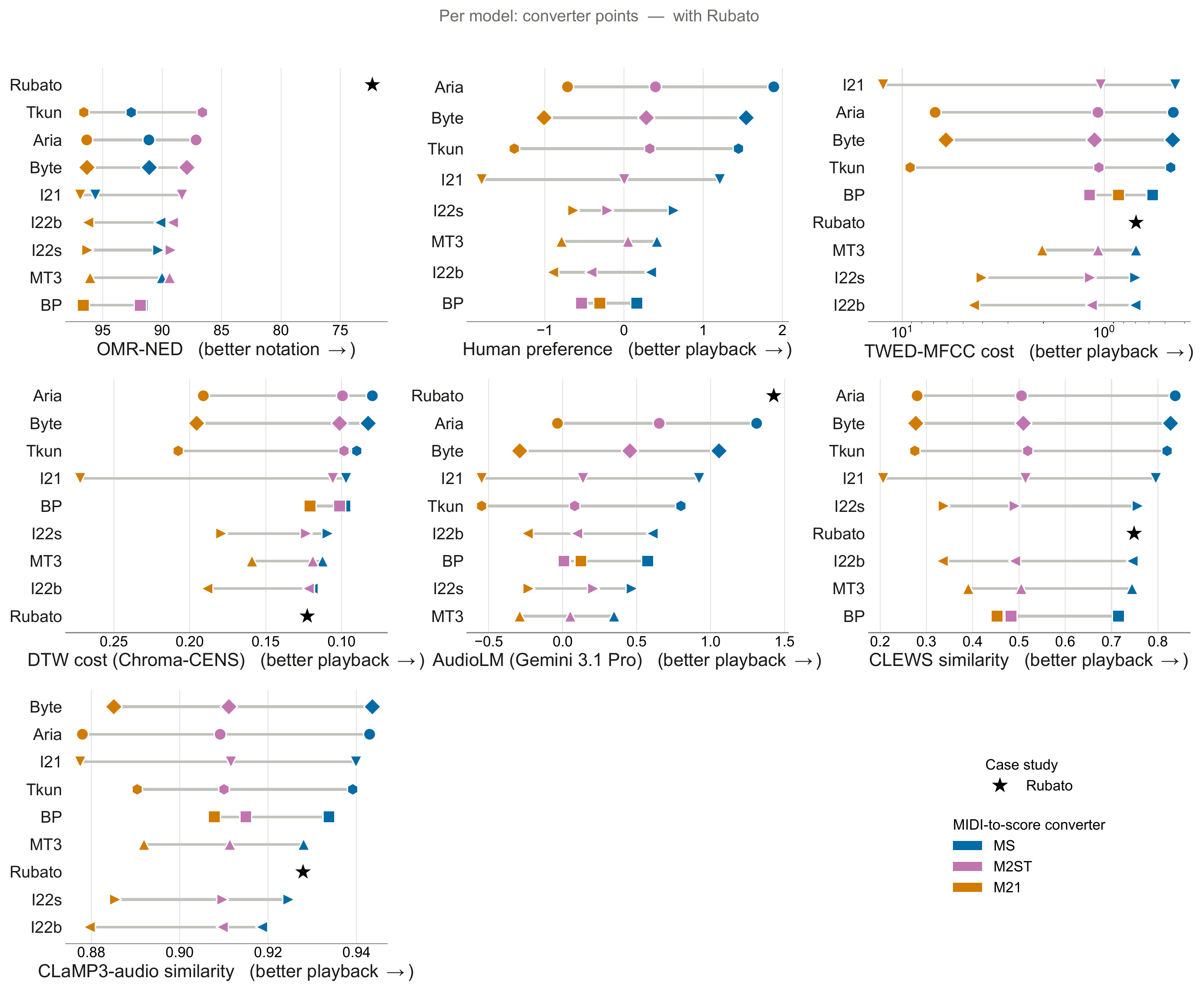}
\caption{Per-model converter ranges for all selected automatic evaluators, with
Rubato shown as a separate end-to-end point. Each row groups the three
converters applied to one audio-to-MIDI model; gray segments show the
minimum--maximum spread, and all axes are oriented so that better results lie to
the right. These plots complement the focused OMR-NED and CLEWS range figures of the
main paper.}
\label{fig:appendix-system-ranges-rubato}
\end{figure*}

\section{Supplementary Cross-Dimension Analysis}
\label{app:omr-cross-dimension-complete}

Figure~\ref{fig:appendix-omr-cross-dimension-grid} shows the complete cross-dimension comparison between OMR-NED and all playback-side or preference-based views considered in the main analysis. This supplements the notation--playback tradeoff figure of the main paper, which reports only the selected OMR-NED versus CLEWS comparison.

\begin{figure*}[p]
\centering
\setlength{\tabcolsep}{2pt}
\begin{tabular}{cc}
\begin{minipage}{0.48\linewidth}
\centering
\includegraphics[width=\linewidth]{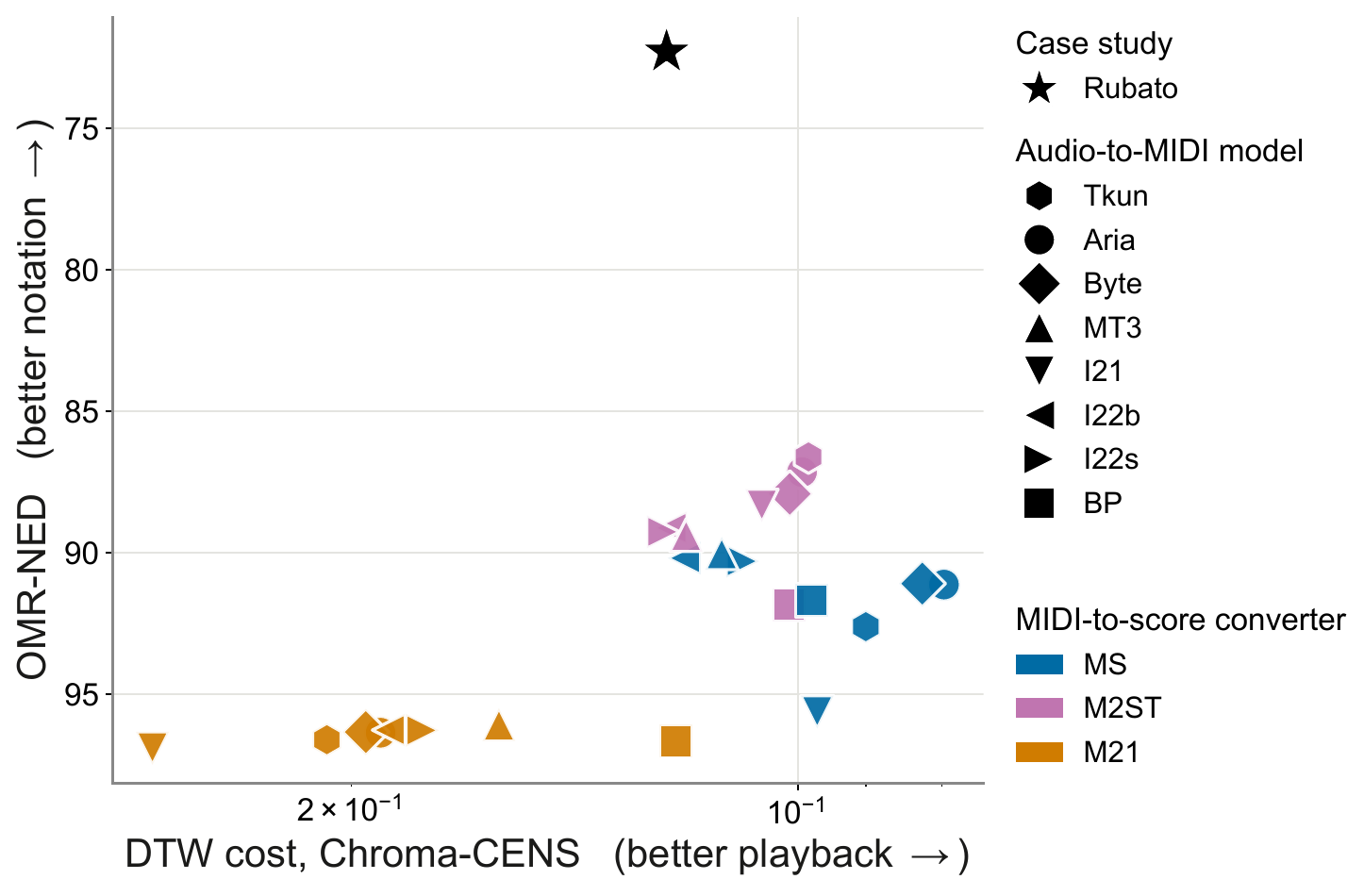}
{\footnotesize (a) OMR-NED vs. DTW--Chroma CENS}
\end{minipage}
&
\begin{minipage}{0.48\linewidth}
\centering
\includegraphics[width=\linewidth]{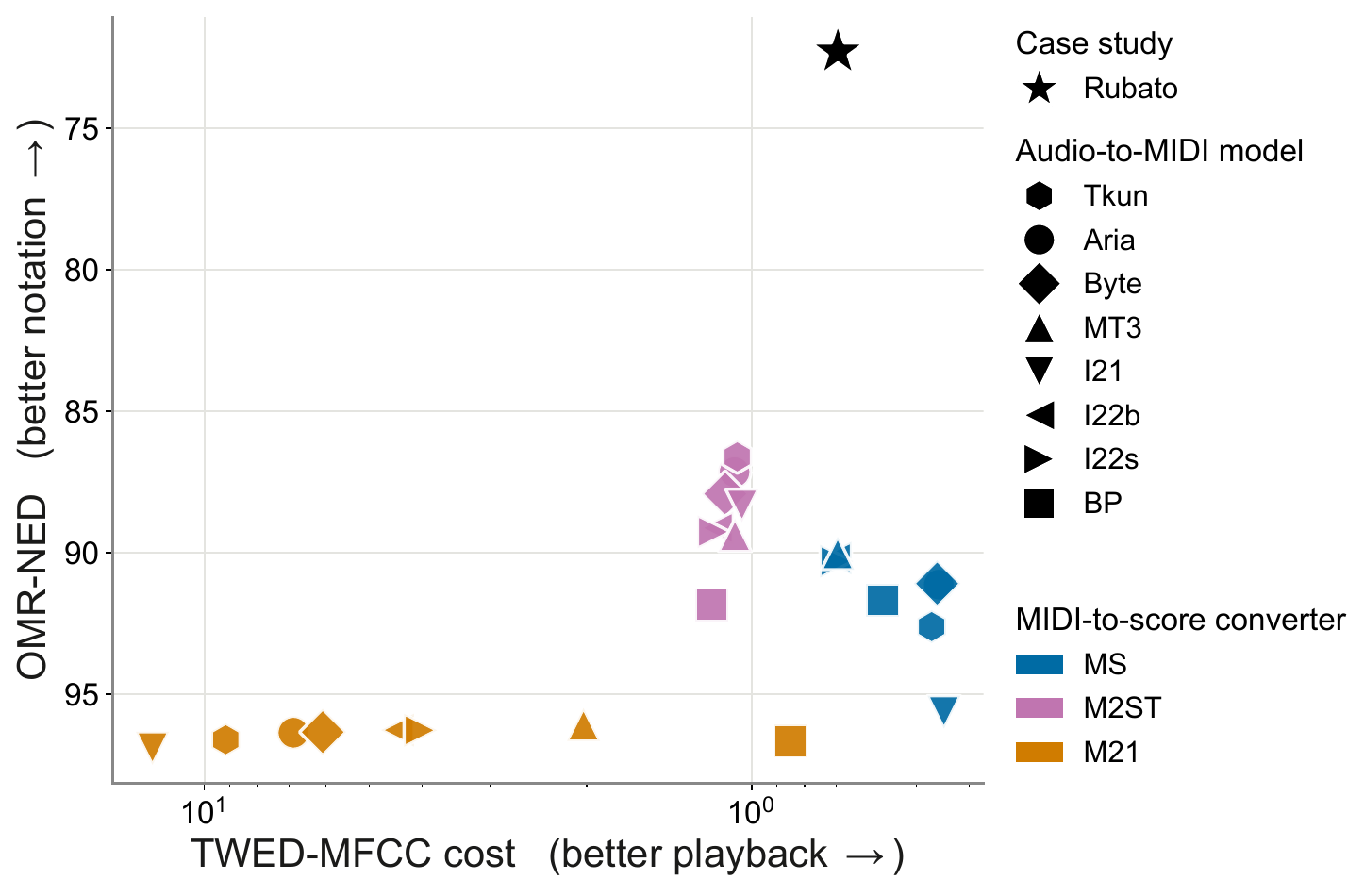}
{\footnotesize (b) OMR-NED vs. TWED--MFCC}
\end{minipage}
\\[0.6em]
\begin{minipage}{0.48\linewidth}
\centering
\includegraphics[width=\linewidth]{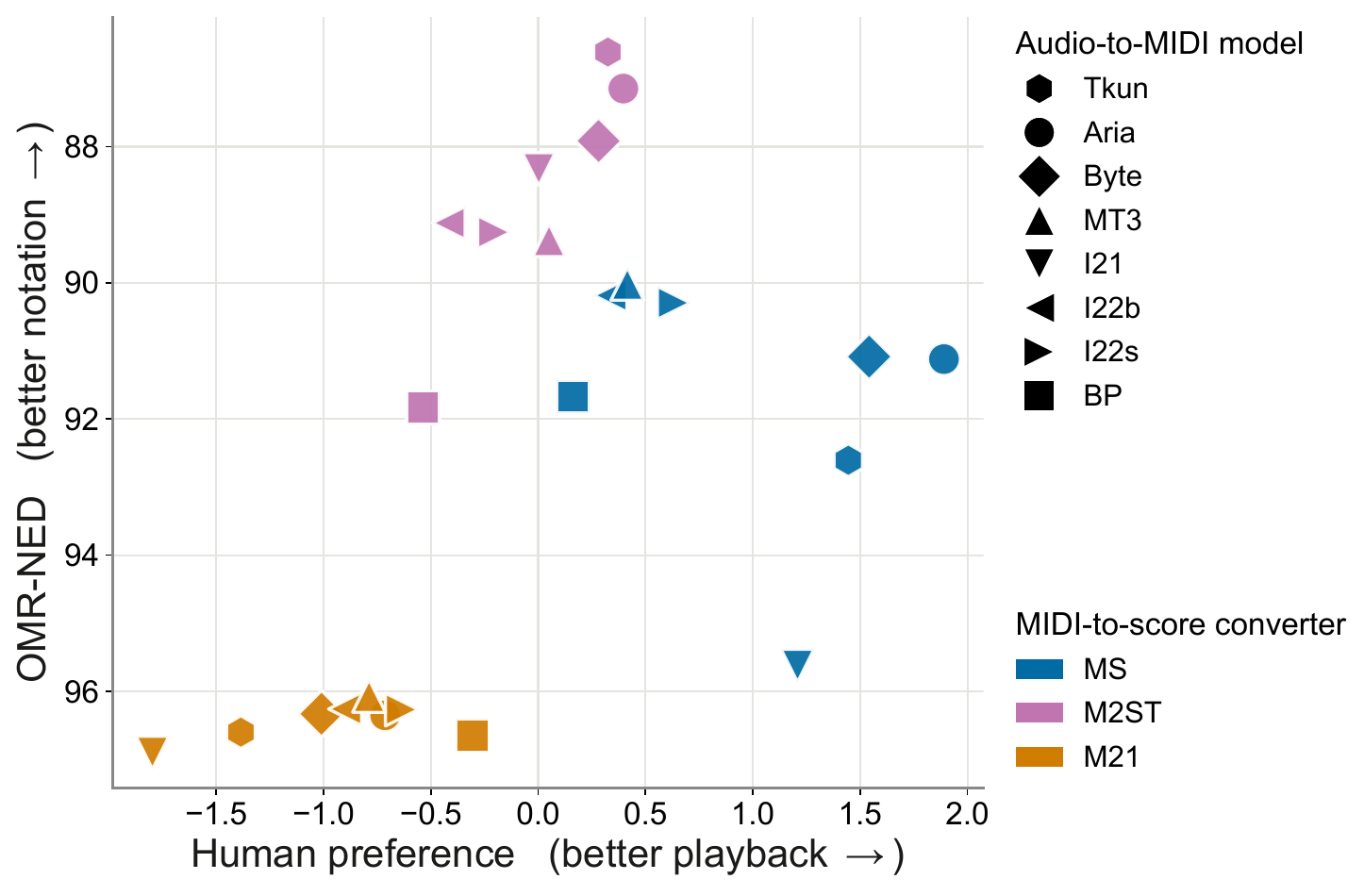}
{\footnotesize (c) OMR-NED vs. human preference}
\end{minipage}
&
\begin{minipage}{0.48\linewidth}
\centering
\includegraphics[width=\linewidth]{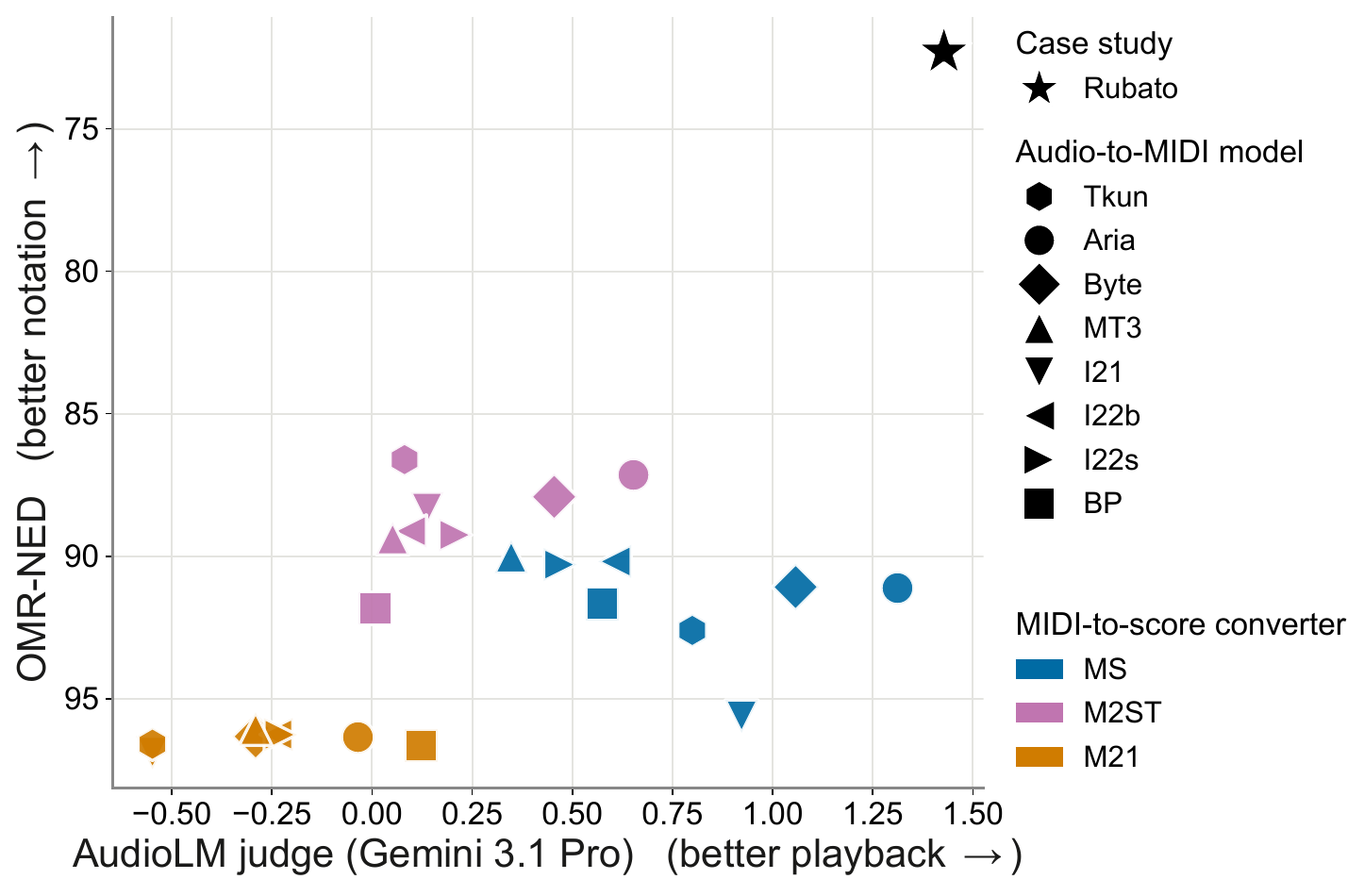}
{\footnotesize (d) OMR-NED vs. Gemini 3.1 Pro}
\end{minipage}
\\[0.6em]
\begin{minipage}{0.48\linewidth}
\centering
\includegraphics[width=\linewidth]{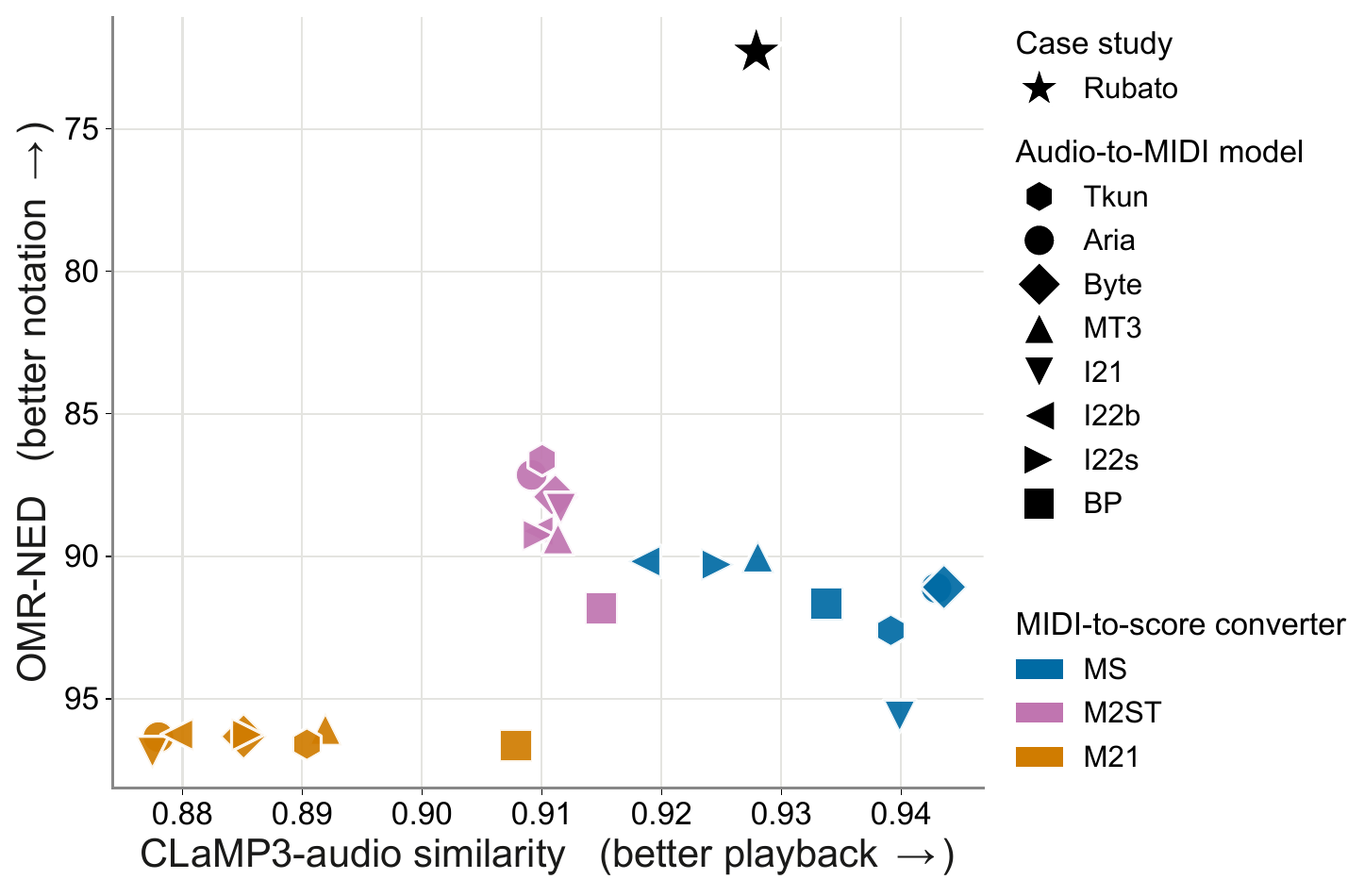}
{\footnotesize (e) OMR-NED vs. CLaMP~3 audio}
\end{minipage}
&
\begin{minipage}{0.48\linewidth}
\centering
\includegraphics[width=\linewidth]{figures/omr_vs_clews.pdf}
{\footnotesize (f) OMR-NED vs. CLEWS}
\end{minipage}
\end{tabular}

\vspace{0.4em}
\caption{Complete cross-dimension comparisons between OMR-NED and playback-side or preference-based evaluation scores over the 24 modular systems. OMR-NED, DTW, and TWED are lower-is-better costs, while human preference, CLaMP~3 audio, and CLEWS scores are higher-is-better similarity or preference scores.}
\label{fig:appendix-omr-cross-dimension-grid}
\end{figure*}

\clearpage
\endgroup